\documentclass{cernrep}
\usepackage{tikz}
\begin{document}
\newcommand{\eqn}[3]{\parbox{9cm}{#1}\hspace{1cm}\parbox{5cm}{\raggedright\bf #2}\hfill\parbox[b]{8mm}{\begin{equation} \label{#3} \end{equation}}}
\title{Cavity types}
\author{F. Gerigk}
\institute{CERN, Geneva, Switzerland}
\maketitle

\begin{abstract}
In the field of particle accelerators the most common use of RF cavities is to increase the particle velocity of traversing particles. This feature makes them one of the core ingredients of every accelerator, and in the case of linear accelerators they are even the dominant machine component. Since there are many different types of accelerator, RF cavities have been optimized for different purposes and with different abilities, e.g., cavities with fixed or variable RF frequency, cavities for short or long pulses/CW operation, superconducting and normal-conducting cavities. This lecture starts with a brief historical introduction and an explanation on how to get from Maxwell's equations to a simple cavity. Then, cavities will be classified by the type of mode that is employed for acceleration, and an explanation is given as to why certain modes are used in particular cavity types. The lecture will close with a comparison of normal versus superconducting cavities and a few words on the actual power consumption of superconducting cavities.   
\end{abstract}

\section{Introduction}

\subsection{Can we accelerate without cavities?}
The first RF linear accelerator (linac) was proposed and tested by Rolf Wider\"{o}e in 1927/1928. He connected a single drift tube between two grounded electrodes to an RF source, which delivered 25\,keV at a frequency of 1\,MHz. In his experiment he accelerated potassium ions to an energy of 50\,keV and thus demonstrated the principle of RF acceleration. This meant that for the first time the maximum obtainable energy of an accelerator was no longer limited by the electrostatic breakdown voltage of DC machines, which were in operation at the time. With this principle it was now possible to multiply the available voltage of an RF source $V_{RF}$ by the number of gaps $N$ and thus to increase the maximum energy gain $\Delta E$ of a particle with charge $q$ to\\
	\eqn{\[\Delta E=qN_{gap}V_{RF}\mbox{.}\]}{energy gain}{eq:eneg}\\

The principle was soon extended to 30 drift tubes and a 10\,MHz RF source by Sloan and Lawrence in 1931 to accelerate mercury ions to $1.26$\,MeV. 

\begin{figure}[h!]
	\begin{center}
	\begin{tikzpicture}[>=stealth, x=5cm/5,red!70!black,thick]
	\draw[->,black,dashed, line width=1pt](-1.9,0) -- (-0.9,0) (-0.5,0) -- (0.3,0) (0.9,0) -- (1.7,0) (2.5,0) -- (3.3,0) (4.3,0) -- (5,0)node[right]{beam};
	\draw[black](5,-3)--(7,-3)--(7,-1)--(5,-1)--(5,-3);
	\draw[black](6,-2) circle (0.5cm); 
	\draw[black](6,-1.3) node {RF source};
	\draw[red,very thick,samples=100,domain=5.7:6.3] plot (\x, {0.2*sin(600*\x )-2});
	\draw[black] (-2.3,-0.4) -- (-2.3,0.4) -- (-1.7,0.4) -- (-1.7,0.2) arc (90:270:0.2) -- (-1.7,-0.4) -- (-2.3,-0.4);
	\draw[black] (-2.5,0.6) node {ion source}; 
	\draw[black] (5,-2.5) -- (-2,-2.5) (5,-1.5) -- (-1,-1.5) (-2,-0.4) -- (-2,-4) (-2.2,-4) -- (-1.8,-4);
	\fill [black] (-2,-2.5) circle (2pt);
	\foreach \x in {0.3,3.5}
		{\fill [black] (\x,-2.5) circle (2pt);
		\draw[black] (\x,-2.5) -- (\x,-0.8);}
	\foreach \x in {-1,1.8}
		{\fill [black] (\x,-1.5) circle (2pt);
		\draw[black] (\x,-1.5) -- (\x,-0.8);}
	\foreach \x in {-1.2,0,1.4,3}
		\draw[black] (\x,0) ellipse (0.3cm and 0.8cm);
	\foreach \x in {-0.8,0.6,2.2,4}
		\draw[black] (\x,-0.8) arc (-90:90:0.3cm and 0.8cm);
	\draw[black] (-1.2,0.8) -- (-0.8,0.8) node [above,midway] {1} (-1.2,-0.8) -- (-0.8,-0.8);
	\draw[black] (0,0.8) -- (0.6,0.8) node [above,midway] {2} (0,-0.8) -- (0.6,-0.8);
	\draw[black] (1.4,0.8) -- (2.2,0.8) node [above,midway] {$\ldots$} (1.4,-0.8) -- (2.2,-0.8);
	\draw[black] (3,0.8) -- (4,0.8) node [above,midway] {n} (3,-0.8) -- (4,-0.8);
	\draw[<->,black] (2.6,1.4) -- (4.2,1.4) node [above, midway] {$l_n$};
	\foreach \x in {-0.5,2.5}
		\draw[->,red, line width=6pt] (\x,0) -- (\x+0.8,0);
	\foreach \x in {-1.8,0.8}
		\draw[<-,red, line width=6pt] (\x,0) -- (\x+0.8,0);
	\end{tikzpicture}
	\caption{Principle of Wider\"{o}e's RF linear accelerator: RF electric fields between drift tubes accelerate the particles}
	\label{fig:wideroe}
	\end{center}
	\end{figure}
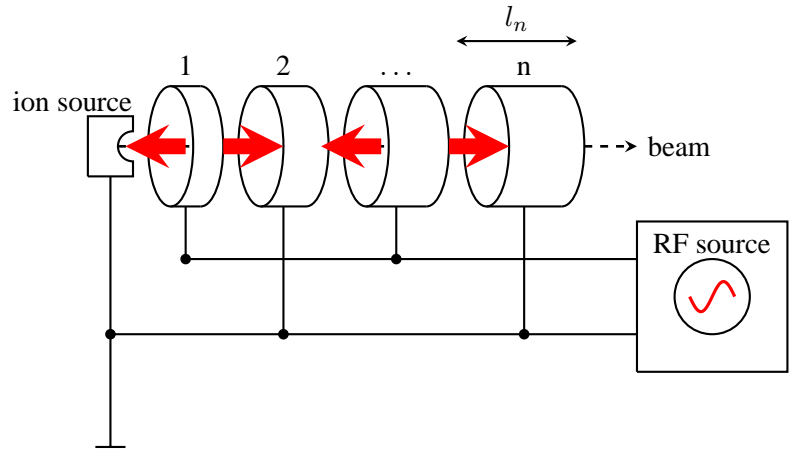

The basic idea of the Wider\"{o}e's linac is shown in Fig.~\ref{fig:wideroe}. Subsequent drift tubes are connected to opposite polarities of an RF source. This means that at a `frozen' point in time every second gap has an RF phase suitable for acceleration. The idea is that while the particles move from one gap to the next the polarity of the RF source changes, so that particles always see an accelerating field in the gaps. In order for this to work, the RF oscillations have to be in synchronism with the passage of the accelerated particles. For a fixed RF frequency $f$ this leads to the following condition for the distance $l_{\,n}$ between the gaps of a Wider\"{o}e's linac:\\
	\eqn{\[l_n=\frac{v}{2 f}\mbox{,}\]}{synchronism condition}{eq:sync}\\
	
\noindent which means that the distance between the gap centres $l_{\,n}$ has to become larger with increasing particle velocity $v$. 

\begin{figure}[h!]
	\begin{center}
	\includegraphics[width=8cm]{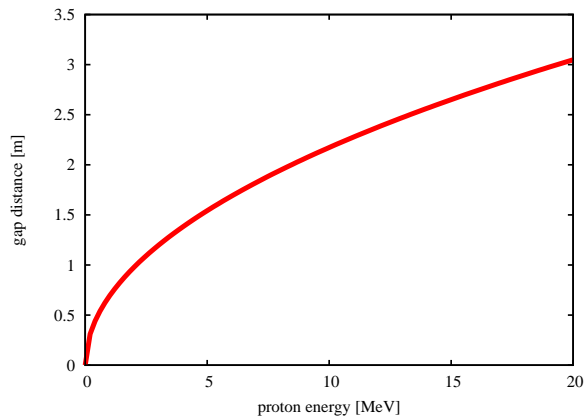}
	\caption{Gap distance for proton acceleration with a 10\,MHz RF source}
	\label{fig:dtprotons}
	\end{center}
	\end{figure}

Even though the Wider\"{o}e linac revolutionized the acceleration of particles, its maximum energy was still severely limited for two reasons: i) For higher velocities the length of the drift tubes has to increase, which means that there was a natural `practical' limit for these machines. Especially for light ions or protons the drift tubes would simply become too long (see Fig.~\ref{fig:dtprotons}). One can solve this problem by going to higher frequencies, but ii) since the accelerating structure is not enclosed by an electric boundary, the operation at higher frequencies ($> 10$\,MHz) meant that the drift tubes were basically becoming antennas. With increasing frequency they radiate more and more of the RF energy instead of using it for acceleration, thus leading to a very poor efficiency of the accelerator. 

\begin{figure}[h!]
	\begin{center}
	\begin{tikzpicture}[>=stealth, x=5cm/5,red!70!black,thick]
	\draw[->,black,dashed, line width=1pt](-3.3,0) -- (-0.9,0) (-0.5,0) -- (0.3,0) (0.9,0) -- (1.7,0) (2.5,0) -- (3.3,0) (4.3,0) -- (6,0)node[right]{beam};
	\draw[black](5.5,-3)--(7.5,-3)--(7.5,-1)--(5.5,-1)--(5.5,-3);
	\draw[black](6.5,-2) circle (0.5cm); 
	\draw[black](6.5,-1.3) node {RF source};
	\draw[red,very thick,samples=100,domain=6.2:6.8] plot (\x, {0.2*sin(600*(\x-0.5) )-2});
	\draw[black] (-3.7,-0.4) -- (-3.7,0.4) -- (-3.1,0.4) -- (-3.1,0.2) arc (90:270:0.2) -- (-3.1,-0.4) -- (-3.7,-0.4);
	\draw[black] (-3.4,0.6) node {ion source}; 
	\draw[black] (5.5,-2.5) -- (1.3,-2.5) -- (1.3,-1.9) arc (0:180:0.2) -- (1.0,-2) (5.5,-1.5) -- (5,-1.5);
	\fill [black] (0.9,-2) circle (2pt);
	\fill [black] (5,-1.5) circle (2pt);
	\draw[black, very thick] (-2,2) -- (4.5,2) (4.5,-2) -- (1.4,-2) (1.2,-2) -- (-2,-2); 
	\draw[black, very thick] (-2,0) ellipse (0.75cm and 2cm);
	\draw[black, very thick] (4.5,-2) arc (-90:90:0.75cm and 2cm);
	\draw[black, very thick] (-2,0) ellipse (0.05cm and 0.133cm);
	\foreach \x in {-1.2,0,1.4,3}
		\draw[gray, very thick] (\x,0) ellipse (0.3cm and 0.8cm);
	\foreach \x in {-0.8,0.6,2.2,4}
		\draw[gray, very thick] (\x,-0.8) arc (-90:90:0.2cm and 0.8cm);
	\draw[gray, very thick] (-1.2,0.8) -- (-0.8,0.8) node [black, above,midway] {1} (-1.2,-0.8) -- (-0.8,-0.8);
	\draw[gray, very thick] (0,0.8) -- (0.6,0.8) node [black, above,midway] {2} (0,-0.8) -- (0.6,-0.8);
	\draw[gray, very thick] (1.4,0.8) -- (2.2,0.8) node [black, above,midway] {$\ldots$} (1.4,-0.8) -- (2.2,-0.8);
	\draw[gray, very thick] (3,0.8) -- (4,0.8) node [black, above,midway] {n} (3,-0.8) -- (4,-0.8);
	\draw[<->,black] (2.6,1.4) -- (4.2,1.4) node [black, above, midway] {$l_n$};
	\foreach \x in {-1.8,-0.6,0.8,2.4,4.2}
		\draw[->,red, line width=6pt] (\x,0) -- (\x+0.8,0);
	\end{tikzpicture}
	\caption{Principle and field profile of an Alvarez linac}
	\label{fig:Alvarez}
	\end{center}
	\end{figure}
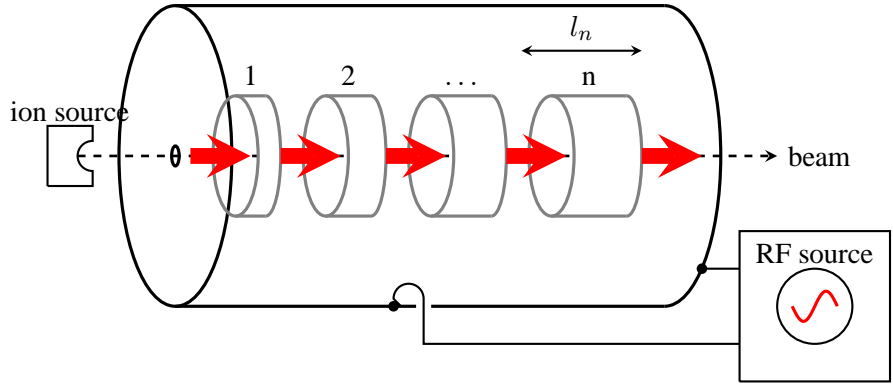

So the answer to the title of this section is ``Yes, we can accelerate without cavities, but reaching higher energies becomes very inefficient''. A solution was finally proposed by Louis Alvarez in 1946, who put the Wider\"{o}e linac into a conducting cylinder. This solved the problem of radiated energy but in addition to the synchronism condition of Eq.~(\ref{eq:sync}) one also has to make sure that each gap has the same inherent resonant frequency, which is now determined by the diameter of the cylinder, the distance between the drift tubes, and their diameter. Figure~\ref{fig:Alvarez} shows the principle of the Alvarez linac
and we can note an important difference to the Wider\"{o}e linac. While for the latter the RF phase of the electric fields changes from gap to gap by $\pi$, in the Alvarez linac the field points in all gaps into the same direction. This comes from the fact that the field direction is given by the lowest (in frequency) resonance, which exists in the surrounding cylinder, and it turns out that for this mode the electric field points in the same direction in all gaps. One speaks of a `zero-mode' because of the zero phase difference between the gaps (more on this later). Here synchronism with the RF demands that the RF phase changes by $2\pi$ while the particles travel from one gap to the next.

The Alvarez linac became possible by the development of high-power, high-frequency RF amplifiers. A technology that had just been invented during World War II to power radar systems. The typical frequency of these early radars was 200\,MHz and by coincidence this frequency results in reasonably sized cylindrical resonators with a diameter of around 1\,m. It is thus not by chance that most early linacs operate at a frequency of 200\,MHz.

\subsection{Classification of cavities according to their application}
\subsubsection{Constant velocity, constant RF frequency}
\label{sec:cvcrf}
The simplest case is to accelerate constant-velocity particles. Since we usually try to accelerate particles we can only speak of `constant velocity' once we have `relativistic' particles, meaning velocities close to the speed of light with a relativistic $\beta$ ($v=\beta c$, with $c$ being the velocity of light) close to one (e.g. $\beta > 0.99$). In this case there is no longer a need to adapt the RF frequency of a synchrotron to the revolution frequency of the particles. In linear accelerators, which have by definition a fixed RF frequency a high $\beta$ means that the distance between accelerating gaps is no longer changing so that one no longer has to adapt the design of the accelerating structure to the increasing particle speed. For constant frequency and $v=c$ we generally find the cavities with the highest accelerating fields at the highest efficiencies. We can find these conditions in electron accelerators (linacs and circular accelerators) already for energies below 1\,MeV, in high-energy proton accelerators above several GeV, and in ion accelerators for energies above 10s or even 100s of GeV (depending on the ion species). 

\subsubsection{Changing velocity, constant RF frequency}
As in the previous case the advantage of having a fixed RF frequency lies in a simple RF system. Only one type of RF power source is needed, and also the RF distribution system can be standardized. These conditions are met in cyclotrons and in low-$\beta$ ion and proton linacs, where the velocity of the particles is increased along the acceleration cycle but where the RF frequency can be kept constant. In cyclotrons, one typically uses single-cell cavities, avoiding any problems with synchronicity between subsequent gaps of multi-cell structures. In linacs, however, where one aims to use multi-cell structures in order to keep the overall linac length compact, it is necessary to adapt the gap distance to the velocity of the particles. This condition leads to a large variety of cavities, which are each optimized for maximum efficiency in their respective velocity ranges. 

\subsubsection{Changing velocity, variable RF frequency}
\label{sec:velvar}
These are the most demanding operating conditions for RF cavities, which have to be met in low-$\beta$ (ion/proton) synchrotrons and FFAGs. A change in RF frequency can be achieved by putting a material with an adjustable permeability (typically ferrites) into a cavity. Owing to the high RF losses only relatively small accelerating voltages (10s of keV) can be achieved in these cavities at rather poor efficiencies. More details can be found in Ref.~\cite{Klingbeil}.

\subsubsection{Non-accelerating cavities}
Technically these cavities should be listed under Section~\ref{sec:cvcrf} but since they are not used for acceleration they deserve a special mention. The two main applications of non-accelerating cavities are i) RF deflection, and ii) RF bunching. RF deflection is used for instance for low-energy beam chopping systems, where a transversely deflecting RF field removes parts of the regular bunch train. One such cavity is used for instance at JPARC \cite{JPARCchopper} and it is designed with a low quality factor in order to minimize the rise and fall time to $\approx 10$\,ns. In the case of the JPARC linac, the bunches arrive with a distance of $\approx 3$\,ns, which means that around three bunches are lost during the rise and fall time, respectively. Since the beam duty cycle is not too high ($1.25$\%) these losses are acceptable, while machines at higher intensity typically use faster stripline choppers \cite{Fritzchopper} in order to reduce the rise and fall time. Another application of RF deflection are CRAB cavities \cite{CRAB}, which are proposed to maximize the collision rate of colliding particle beams, which are brought to collision under a certain angle. The third application for RF deflection is the combination of two beams arriving at a certain angle into one common beam line, which is often referred to as funnelling. In the most demanding funnelling mode bunches are interleaved one by one, which means that the bunch repetition frequency doubles during this process. 

Another application of non-accelerating cavities are bunching cavities, which are used to keep bunches longitudinally confined during beam transport. In early accelerators bunchers were also used to actually form bunches out of CW beams coming from a particle source. In modern linacs the job of bunch-formation (and acceleration) is done by Radio Frequency Quadrupoles (RFQs).

Finally one should also mention the use of RF cavities for beam monitoring, which is covered by Braun in lecture \cite{HBraun}.

\section{From Maxwell to RF cavities}
The basics of RF are the subject of several lectures of this school so we will review only the main equations, which are important to understand the principle of RF cavities. The starting point for all RF problems are Maxwells's equation, which we quote here in their differential form.\\
	\parbox{9cm}{\begin{align*}
	\nabla \times \mathbf{H} &= \mathbf{J} + \frac{\partial\mathbf{D}}{\partial t}\\
	\nabla \times \mathbf{E} &= -\frac{\partial\mathbf{B}}{\partial t} \\
	\nabla \cdot \mathbf{D} &= q_v \\
	\nabla \cdot \mathbf{B} &= 0
	\end{align*}}\hspace{1.0cm}\parbox{5cm}{\raggedright\bf Maxwell's equations}\hfill
	\parbox[b]{8mm}{\begin{equation}\label{eq:maxw}\end{equation}}

These equations can be solved for instance in cylindrical coordinates ($r$, $\varphi$, $z$) for the electric and magnetic field components. The simplest solution having an axial electric field is the so-called TM$_{01}$ wave (the nomenclature of waves/modes will be explained in Section~\ref{sec:Nomenclature}), which consists of radial and longitudinal electric field components and an azimuthal magnetic field component. All other field components are equal to zero. The fields of a TM$_{\,01}$ wave travelling in positive $z$-direction  are\\
 	\parbox{9cm}{\begin{align*}
	E_r &= j\frac{k_z}{k_c} E_0 J_1 \left( k_c r\right) e^{-j k_z z } e^{j\omega t}\mbox{,} \\
	E_z &= E_0 J_0 \left(k_c r \right) e^{-jk_z z} e^{j\omega t}\mbox{,} \\
	H_{\varphi} &= j\frac{k}{Z_0 k_c}E_0 J_1 \left( k_c r \right) e^{-j k_z z} e^{j\omega t}\mbox{,}
	\end{align*}}\hspace{1.0cm}\parbox{5cm}{\raggedright\bf TM{\boldmath$_{01}$} mode in cylindrical coordinates}\hfill
	\parbox[b]{8mm}{\begin{equation}\label{eq:TM01}\end{equation}}

\noindent where $E_0$ is the amplitude of the electric field, $j$ indicates imaginary parts, $J_{\,0}$ and $J_1$ are Bessel functions of the first kind of zeroth and first order, $\omega= 2\pi f$ is the angular frequency, $k$ is the wave number, $k_{\,z}$ the wave propagation constant, and $Z_0$ is the wave-impedance in free space:\\
 	\parbox{9cm}{\begin{align*}
	k &= \frac{2\pi}{\lambda} = \frac{\omega}{c}\mbox{,} \\
	Z_0 &= \sqrt{\frac{\mu_0}{\varepsilon_0}} = 377\,\Omega\mbox{.}
	\end{align*}}\hspace{1.0cm}\parbox{5cm}{\raggedright $\begin{array}{l} \mbox{\bf wave number} \\ \\ \mbox{\bf free-space impedance} \end{array}$}\hfill
	\parbox[b]{8mm}{\begin{equation} \end{equation}}

In the form in which Eq.~(\ref{eq:TM01}) is written we have already anticipated a cylindrical boundary condition, which we assume as a perfectly conducting pipe with the radius $a$. This boundary forces $E_{\,z}=0$ at $r=a$ (which means $J_0(k_c a) = 0$) and leads to a certain cut-off wave-length $\lambda_{\,c}$ below which, all waves will be damped exponentially. Only waves with shorter wave-length, or a frequency above the cut-off frequency $\omega_{c}$ can propagate in the pipe undamped. The relations between cut-off wave-length, cut-off wave-number $k_c$, and the propagation constant $k_z$ of the wave are:\\
 	\parbox{9cm}{\begin{align*}
	\lambda_c &\approx 2.61 a\mbox{,} \\ 
	k_z^2 &= k^2 - k_c^2\mbox{,} \\
	k_c &= \frac{2\pi}{\lambda_c} = \frac{\omega_c}{c}\mbox{.}
	\end{align*}}\hspace{1cm}\parbox{5cm}{\raggedright {\bf cut-off wave length (TM{\boldmath$_{01}$})} \\ \vspace{0.2cm} {\bf propagation constant}  \\ \vspace{0.2cm} {\bf cut-off wave-number}}\hfill
	\parbox[b]{8mm}{\begin{equation} \label{eq:cutoff} \end{equation}}
	
\begin{figure}[ht]
	\begin{center}	
	\begin{tikzpicture}[>=stealth, x=5cm/5,red!70!black,thick]
	\draw[->,black](-5,0)  -- (5,0)node[right]{$k_{z}$};
	\draw[->,black](0,0)node[below]{$0$}  -- (0,5)node[above]{$\omega$};
	\pgfplothandlerlineto
	\pgfplotfunction{\x}{0,0.01,...,5}{\pgfpointxy{\x}{sqrt(1+(\x^2))}}
	\pgfusepath{stroke}
	\pgfplothandlerlineto
	\pgfplotfunction{\x}{0,0.01,...,5}{\pgfpointxy{-\x}{sqrt(1+(\x^2))}}
	\pgfusepath{stroke}
	\draw [->](1.5,2.75)node[above]{$v_{\text{ph}}>c$}--(2,2.25);
	\draw [->](-0.5,1.7)node[above]{$\omega_{c}$}--(0,1);
	\draw[dashed,blue!70!black,thick] (0,0) -- (5,5);
	\draw[dashed,blue!70!black,thick] (0,0) -- (-5,5);
	\draw [->,blue!70!black](3.5,2.5)node[right]{$v_{\text{ph}}=c$}--(3.01,3.01);
	\draw [black, very thick](1,0)--(1,1.41)--(0,1.41);
	\draw [black](1,0.6)node[right]{$v_{\text{ph}}=\frac{\omega}{k_{z}}$};
	\end{tikzpicture}
	\caption{Brioullin diagram for wave propagation in a cylindrical pipe}
	\label{fig:brioullin}
	\end{center}
	\end{figure}
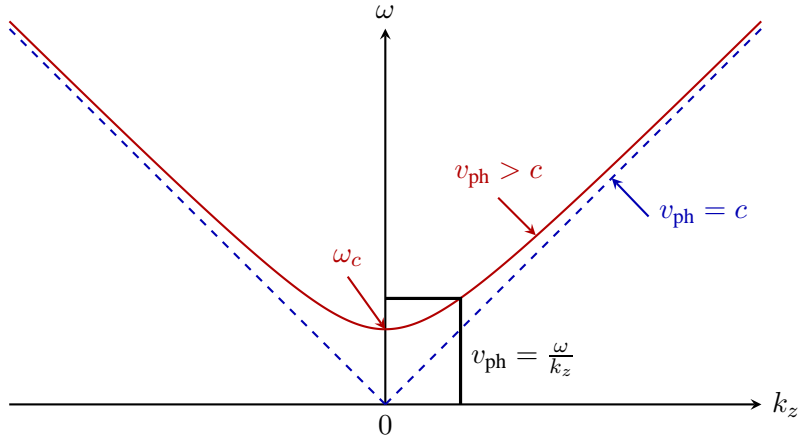

\noindent Using the phase velocity \\
 	\parbox{9cm}{\[
	v_{ph} = \frac{\omega}{k_z}
	\]}\hspace{1.0cm}\parbox{5cm}{\raggedright\bf phase velocity}\hfill
	\parbox[b]{8mm}{\begin{equation} \label{eq:vph} \end{equation}}

\noindent and the propagation constant of Eq.~(\ref{eq:cutoff}) one can establish a dispersion relation between the angular frequency and the propagation constant for our cylindrical pipe:\\
 	\parbox{9cm}{\[
	k_z^2 = \frac{\omega^2}{v_{ph}^2} = \frac{\omega^2}{c^2} - \frac{\omega_c^2}{c^2}
	\]}\hspace{1.0cm}\parbox{5cm}{\raggedright\bf dispersion relation}\hfill
	\parbox[b]{8mm}{\begin{equation} \label{eq:disprel} \end{equation}}

\noindent which can be plotted in the form of the so-called Brioullin diagram (see Fig.~\ref{fig:brioullin}).

\noindent From the dispersion relation and the Brioullin diagram one can see that
\begin{itemize}
\item each frequency corresponds to a certain phase velocity,
\item the phase velocity $v_{ph}^2 = c^2\frac{\omega^2}{\omega^2-\omega_c^2}$ is always larger than the speed of light,
\item at $\omega = \omega_c$ the propagation constant $k_z$ goes to zero and the phase velocity $v_{ph}=\omega/k_z$ becomes infinite, 
\item it is impossible to accelerate particles in a circular waveguide because synchronism between the particles and the RF is impossible (particles would have to travel faster than light to be synchronous with the RF),
\item information and therefore energy travels at the speed of the group velocity $v_{gr}=$d$\omega/$d$k_z$, and is always slower than the speed of light.
\end{itemize} 

\subsection{Travelling wave cavities}
\begin{figure}[h!]
	\begin{center}
		\begin{tikzpicture}[>=stealth, x=5cm/5,red!70!black,thick]
	\draw[->,black,dashed, line width=1pt](-2,0)  -- (2.5,0)node[right]{beam};
	\foreach \x in {-1.4,-0.7,0,0.7,1.4}
		\draw[black,fill=gray!50] (\x,0) ellipse (0.2cm and 0.8cm);
	\foreach \x in {-1.4,-0.7,0,0.7,1.4}
		\draw[black,fill=white] (\x,0) ellipse (0.12cm and 0.6cm);
	\draw[black] (-1.4,0.8) -- (1.4,0.8) (-1.4,-0.8) -- (1.4,-0.8);
	\draw[<->,black] (-1.4,-1) -- (1.4,-1) node [below,midway] {$L$};
	\draw[<->,black] (-1.8,-0.6) -- (-1.8,0.6) node [left] {$2a$};
	\draw[<->,black] (1.8,-0.8) -- (1.8,0.8) node [right] {$2b$};
	\draw[->,black] (-0.5,1) -- (-0.05,1);
	\draw[->,black] (0.5,1) -- (0.05,1) node [above,midway] {$h$};
	\end{tikzpicture}
	\caption{Simple geometry of a travelling wave structure}
	\label{fig:discs}
	\end{center}
	\end{figure}
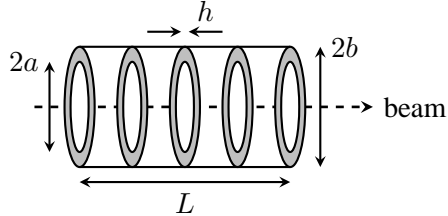

In order to use some kind of cylindric wave-guide to accelerate particles, the phase velocity in the structure needs to be slowed down. This can be achieved by putting some `obstacles' into the wave-guide. For instance some discs as shown in Fig.~\ref{fig:discs}.

Since the electromagnetic wave is `travelling' through the structure, one speaks of `travelling wave' structures in contrast to `standing wave' structures, which will be introduced in the following section. 

In standard accelerator text books (e.g., Wangler, see Bibliography) one can find the dispersion relation for a disc-loaded travelling wave structure with the geometry as in Fig.~\ref{fig:discs}:\\
 	\parbox{9cm}{\[
	\omega = \frac{2.405 c}{b}\sqrt{1 + \kappa\left(1-\cos\left(k_z L\right) e^{-\alpha h}\right)}
	\]}\hspace{1.0cm}\parbox{5cm}{\raggedright\bf dispersion relation for disc-loaded travelling wave structure}\hfill
	\parbox[b]{8mm}{\begin{equation} \label{eq:dispdisc} \end{equation}}

\noindent with $\kappa$ and $\alpha$ (attenuation per unit length) being defined as \\
 	\parbox{10cm}{\[
	\kappa = \frac{4a^3}{3\pi J_1^2 (2.405) b^2 L} \ll 1 \hspace{0.5cm} \mbox{and} \hspace{0.5cm} \alpha \approx \frac{2.405}{a}\mbox{.}
	\]}\hspace{1cm}\parbox{4cm}{\raggedright\bf }\hfill
	\parbox[b]{8mm}{\begin{equation} \label{eq:kappa} \end{equation}}

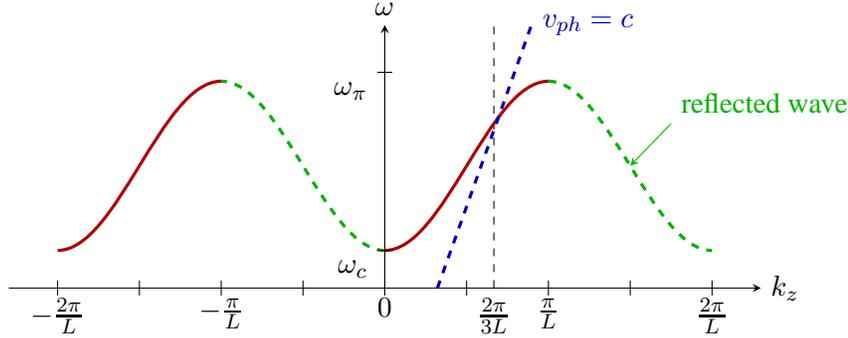
\begin{figure}[h!]
	\begin{center}
	\begin{tikzpicture}[>=stealth]
	\draw[->,black](-5,-0.5)  -- (5,-0.5)node[right]{$k_{z}$};
	\foreach \x in {-4,-3,...,4}
	\draw [x=4*1.57ex](\x,-0.6) -- (\x,-0.4);
	\draw[->,black](0,-0.5)node[below]{$0$}  -- (0,3)node[above]{$\omega$};
	\draw [x=4*1.57ex,dashed](4/3,-0.5)node[below]{$\frac{2\pi}{3L}$}--(4/3,3);
	\draw[x=-4*1.57ex,y=6.5ex,dashed,green!70!black,very thick] (0,0)cos(1,1) sin (2,2);
	\draw[x=-4*1.57ex,y=6.5ex,red!70!black,very thick](2,2)cos (3,1) sin (4,0);
	\draw[x=4*1.57ex,y=6.5ex,red!70!black,very thick] (0,0)cos(1,1) sin (2,2);
	\draw[x=4*1.57ex,y=6.5ex,dashed,green!70!black,very thick](2,2)cos (3,1) sin (4,0);
	\draw [x=4*1.57ex,y=6.5ex,->,green!70!black](3.5,1.5)node[above right]{reflected wave}--(3,1);
	\draw [dashed,blue!70!black,very thick](0.7,-0.5)--(1.95,3)node[right]{$v_{ph}=c$};
	\draw [x=4*1.57ex](-4,-0.5)node[below]{$-\frac{2\pi}{L}$};
	\draw [x=4*1.57ex](4,-0.5)node[below]{$\frac{2\pi}{L}$};
	\draw [x=4*1.57ex](-2,-0.5)node[below]{$-\frac{\pi}{L}$};
	\draw [x=4*1.57ex](2,-0.5)node[below]{$\frac{\pi}{L}$};
	\draw (0.1,0)--(-0.1,0) node[below left]{$\omega_{c}$};
	\draw (0.1,2.37)--(-0.1,2.37) node[below left]{$\omega_{\pi}$};
	\end{tikzpicture}
	\caption{Dispersion diagram for a disc-loaded travelling wave structure. Here the chosen operating point is $v_{ph} = c$ and $k_z = 2\pi/3L$}
	\label{fig:dispersion-discs}
	\end{center}
	\end{figure}

The dispersion relation \eqref{eq:dispdisc} is plotted in Fig.~\ref{fig:dispersion-discs} and one can see that there are indeed modes (in this example: $2\pi/3 < k_{\,z} L < \pi$) with a phase velocity equal to or even lower than the speed of light. It should be noted that for different geometries it is possible to have a different range of modes. When a structure operates in the $2\pi/3$ mode it means that the RF phase shifts by $2\pi/3$ per cell, or in other words one RF period stretches over three cells. The particles then move in synchronism with the RF phase from cell to cell.  

By attaching an input and output coupler to the outermost cells of the structure we obtain a usable accelerating structure. Since the particles gain energy in every cell, the electromagnetic wave becomes more and more damped along the structure, extracted via the output coupler, and then dumped in an RF load. If one is interested to have the maximum possible accelerating gradient in each cell, then one can counteract the decreasing fields by changing the bore radius from cell to cell. The idea is to slow down the group velocity from cell to cell and to obtain a `constant gradient' structure, rather than a `constant impedance' structure where the bore radii are kept constant. Other optimizations, e.g., for maximum efficiency are also possible.

\subsection{Standing wave cavities}
One obtains a cylindrical standing wave structure by simply closing both ends of a circular wave guide with electric walls. This will yield multiple reflections on the end walls until a standing wave pattern is established. Owing to the additional boundary conditions in the longitudinal direction, we get another `restriction' for the existence of electromagnetic modes in the structure. While the longitudinally open travelling wave structure allows all frequencies and cell-to-cell phase variations on the dispersion curve, now only certain `loss-free' modes (still assuming perfectly conducting walls) with discrete frequencies and discrete phase changes can exist in the cavity. If one feeds RF power at a different frequency, then the excited fields will be damped exponentially (for the experts: strictly speaking this statement is only true if we assume zero bandwidth of the modes), similar to the modes below the cut-off frequency of a wave-guide. 

The corresponding dispersion relation for standing wave cavities can again be found in textbooks (Wangler or also Ref.~\cite{CASMaurizio}). However, one should pay attention as to whether the structure under consideration has magnetic cell-to-cell coupling or electric cell-to-cell coupling and which kind of end-cell is used for the study. The most common form of the dispersion relation is derived from a coupled circuit model with $N+1$ cells. Usually the model has half-cell terminations on both sides of the chain, which practically represents the behaviour of an infinite chain of electrically coupled resonators (compare with the original papers by Nagle and Knapp \cite{Knapp}):\\
 	\eqn{\[\omega_n = \frac{\omega_{0}}{\sqrt{1+k\cos\left(n\pi/N \right)}} \mbox{, }n=0,1,\ldots ,N\mbox{.}\]}{dispersion relation for half-cell terminated standing wave structure}{eq:dispstandmag} 

\noindent Assuming an uneven number of cells, $\omega_{\,0}$ is the frequency of the $\pi/2$ mode. A more general definition is to say that $\omega_{0}$ is the frequency of the uncoupled single cells. $k$ is the cell-to-cell coupling constant, and $n\pi/N$ the phase shift from cell to cell. For $k\ll 1$, which is usually fulfilled, the coupling constant is given by\\
 	\eqn{\[k = \frac{\omega_{\pi -mode}-\omega_{0-mode}}{\omega_{0}}\mbox{.}
	\]}{coupling constant}{eq:couplconstant} 

Two characteristics of the dispersion curve are worth noting: i) the total frequency band of the mode $\omega_{\pi - mode}-\omega_{0-mode}$ is independent of the number of cells, which means that we can determine the cell-to-cell coupling constant by measuring the complete structure (only true if all coupling constants are equal); ii) for electric coupling the 0-mode has the lowest frequency and the $\pi$-mode has the highest frequency. In case of magnetic coupling this behaviour is reversed and one can find the corresponding dispersion curve by changing the sign before the coupling constant in Eq.~\eqref{eq:dispstandmag}.

\begin{figure}[h!]
	\begin{center}
	\begin{tikzpicture}[domain=0:7,samples=60]
	\draw[red,very thick, mark=x,mark repeat=10] plot (\x, {80/(sqrt(1-0.05*cos(180*\x /7)))});
	\draw[black] (0,77.5) -- (7,77.5);
	\draw[black] (0,77.5) -- (0,82.5);
	\draw[black] (7,77.5) -- (7,82.5);
	\draw[black] (0,77.5) node [below]{0};
	\draw[black] (7,77.5) node [below]{$\pi$};
	\draw[black] (3.5,77.7) -- (3.5, 77.5) node [below]{$\pi/2$};
	\draw[black] (-0.2,82.078) -- (0,82.078) node [left]{$\frac{\displaystyle\omega_0}{\displaystyle\sqrt{1-k}}$};
	\draw[black] (7.2,82.078) -- (7,82.078);
	\draw[black] (-0.2,78.072) -- (0,78.072) node [left]{$\frac{\displaystyle\omega_0}{\displaystyle\sqrt{1+k}}$};
	\draw[black] (7.2,78.072) -- (7,78.072);
	\end{tikzpicture}
	\caption{Dispersion diagram for a standing wave structure with seven magnetically coupled cells}
	\label{fig:dispersion-standing}
	\end{center}
	\end{figure}
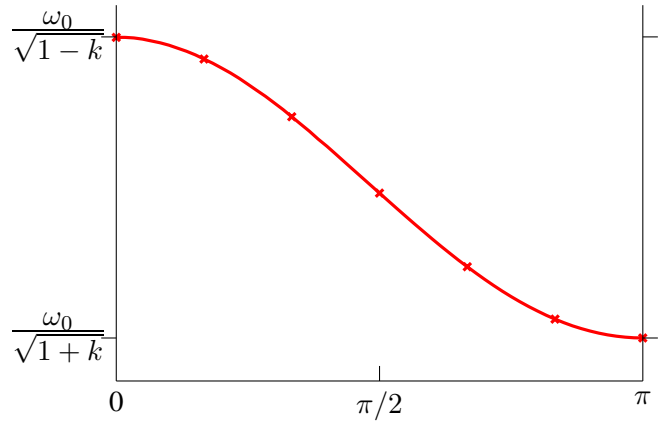

In Fig.~\ref{fig:dispersion-standing} we plot the dispersion curve for a seven-cell (half-cell terminated) magnetically coupled structure according to Eq.~\eqref{eq:dispstandmag}.

In practice one usually has cavities with full-cell termination, and in this case one has to detune the frequency of the end cells to obtain a flat field distribution in the cavity \cite{Schriber01}. In this case one can have a flat field distribution for the $0$-mode or the $\pi$-mode but not for both at the same time because the end cells have to be detuned by a different amount \cite{Schriber2}. 

\subsection{Standing wave versus travelling wave}
The principal difference between the two types of cavity is how and how fast the cavities are filled with RF power. Travelling wave structures are filled `in space', which means that basically cell after cell is filled with power. For the following estimations we consider a frequency range in the 100s of MHz: the filling of a travelling wave structure typically takes place with a speed of approximately 1--3\% of the speed of light and results in total filling times in the sub-microsecond range. Standing wave structures on the other hand are filled `in time': the electromagnetic waves are reflected at the end-walls of the cavity and slowly build up a standing wave pattern at the desired amplitude. In normal-conducting cavities this process is typically in the range of 10s of microseconds. In superconducting cavities the filling process can easily go into the millisecond range (depending on the required field level, the accelerated current, and the cavity parameters). This means that for applications that require very short beam pulses ($< 1\,\mu$s) travelling wave structures are much more power efficient. For longer pulses ($> n\times 10\,\mu$s) both structures can be optimized to similar efficiencies and cost. Since one can have extremely short RF pulses in travelling wave structures, one can obtain much higher peak fields than in standing wave structures. This is demonstrated by the CLIC \cite{clic} accelerating structures, which have reached values of $\approx 100$\,MV/m (limited by breakdown of the electric field), while the design gradient for the superconducting (standing wave) ILC \cite{ilc} cavities is just slightly above 30\,MV/m (generally limited by field emission, and quenches caused by peak magnetic fields). 

Travelling wave structures can theoretically be designed for non-relativistic particles. In existing accelerators, however, they are mostly used for relativistic particles. Low-beta acceleration is typically performed with standing wave cavities.  

For lack of an obvious criterion (other than the pulse length, or particle velocity), one has to do the optimization and cost exercise for each specific application in order to decide which structure is more efficient. Two excellent papers \cite{Miller86, Moiseev00}, which perform this exercise can be used as reference. 

\section{Cavity types classified by electromagnetic modes}
\subsection{TM mode cavities}
\begin{figure}[h!]
	\begin{center}
	\includegraphics[width=4cm]{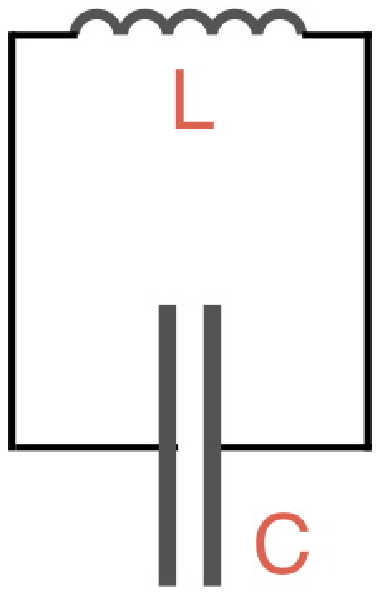}
	\includegraphics[width=4cm]{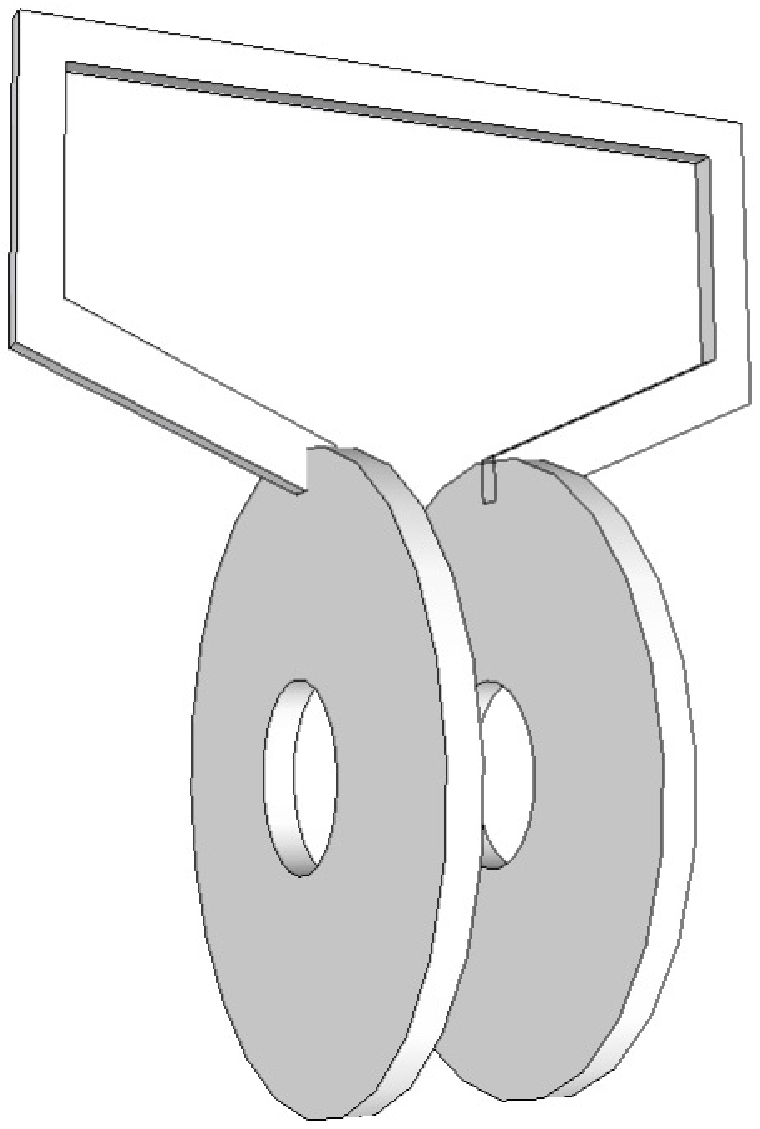}
	\includegraphics[width=4cm]{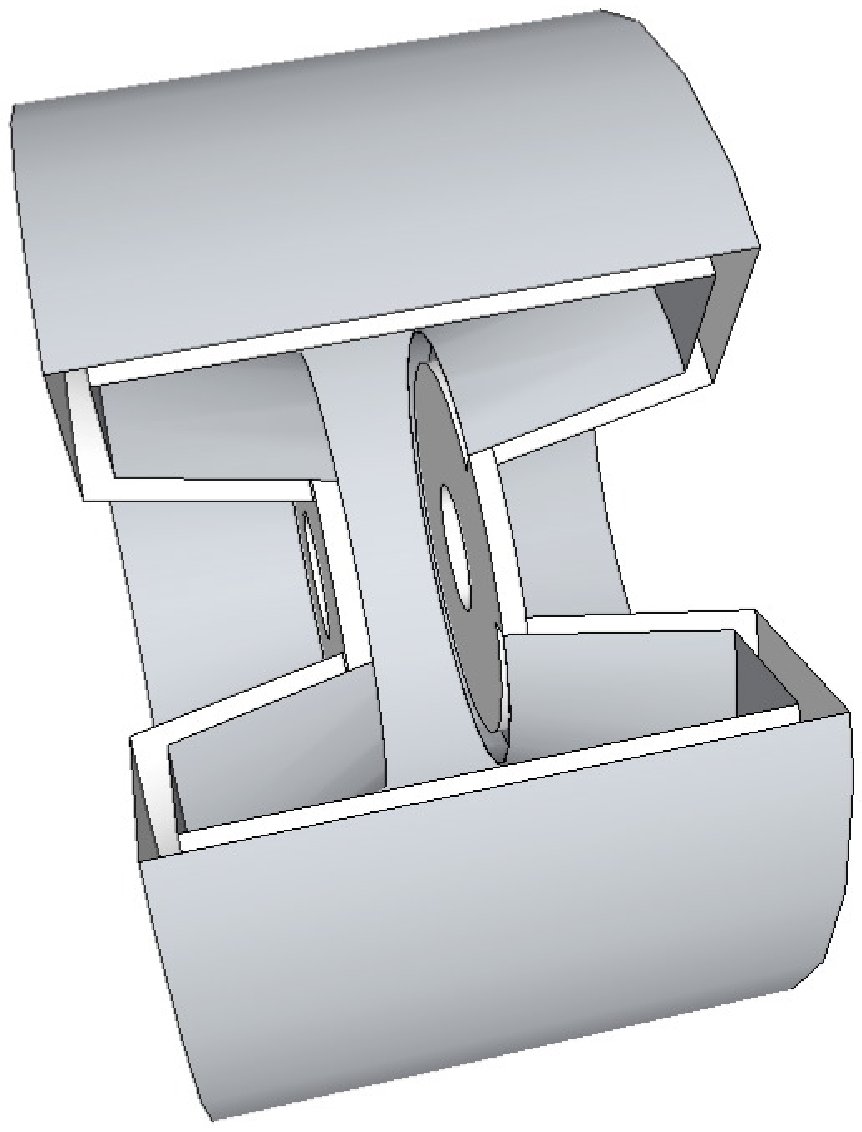}
	\caption{Transition from a lumped resonating circuit to a resonating cavity}
	\label{fig:lumped}
	\end{center}
	\end{figure}

Resonating cavities can be represented conveniently by a lumped element circuit of an inductor (storage of the magnetic energy) and a capacitor (storage of electric energy). Looking at Fig.~\ref{fig:lumped} one can easily imagine how the lumped circuit can be transformed into a cavity.

The simplest case is the so-called pill-box cavity, which is nothing else than an empty cylinder with a conducting surface. The simplest mode in this cavity is the so-called TM$_{\,010}$ mode, which has zero full-period variations in the azimuthal direction ($\varphi$), one `zero' of the axial field component in the radial direction ($r$), and zero half-period variations in the longitudinal ($z$) direction. We can derive its field equations from the TM$_{01}$ field components of the circular wave-guide in Eq.~\eqref{eq:TM01} simply by doing the following:

\begin{itemize}
	\item zero field variation in longitudinal direction means: $k_z=0$ $\Rightarrow$ $k=k_c$;
	\item in standing wave cavities the term `cut-off' frequency does not really make sense, so we replace the expression $k_c$ by $k_r$, indicating that we do have a radial dependency of the axial field component, which can also be interpreted as a radial wave-number;
	\item instead of having only one wave travelling in the positive $z$ direction, we now have one backward and one forward travelling wave, so we replace in the $E$ field terms: $e^{j\omega t}$ by $e^{j\omega t} + e^{-j\omega t} = 2\cos{\omega t}$;
	\item we do the same for the $H$ field term, but here we have to consider that owing to the direction of propagation, the electric and magnetic field are 180$^{\,\circ}$ out of phase, which means that we replace $e^{j\omega t}$ by $e^{j\omega t} - e^{-j\omega t} = 2j\sin{\omega t}$;
\end{itemize}

\noindent then we re-normalize $E_0$ and arrive at\\
 	\parbox{9cm}{\begin{align*}
	E_r &= 0\mbox{,} \\
	E_z &= E_0 J_0 \left(k_r r \right) \cos(\omega t)\mbox{,} \\
	H_{\varphi} &= -\frac{E_0}{Z_0} J_1 \left( k_r r \right) \sin(\omega t)\mbox{.}
	\end{align*}}\hspace{1.0cm}\parbox{5cm}{\raggedright\bf TM{\boldmath$_{010}$} mode in a pill-box cavity}\hfill
	\parbox[b]{8mm}{\begin{equation}\label{eq:TM010}\end{equation}}

Since there is no field dependency on $z$ and $\varphi$, the frequency of the pill-box cavity is only determined by its radius $r=R_{cav}$ and can be written as\\
 	\parbox{9cm}{\[
	\omega_0 = k_r c = \frac{2.405 \cdot c}{R_{cav}}\mbox{.}
	\]}\hspace{1.0cm}\parbox{5cm}{\raggedright\bf angular frequency of pill-box cavity}\hfill
	\parbox[b]{8mm}{\begin{equation} \label{eq:fpillbox} \end{equation}}

\begin{figure}[h!]
	\begin{center}
	\includegraphics[width=2cm]{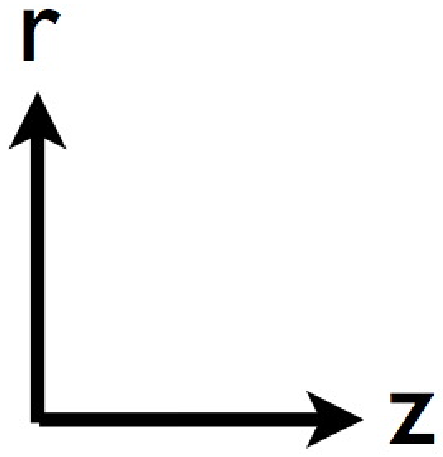}
	\includegraphics[width=4cm]{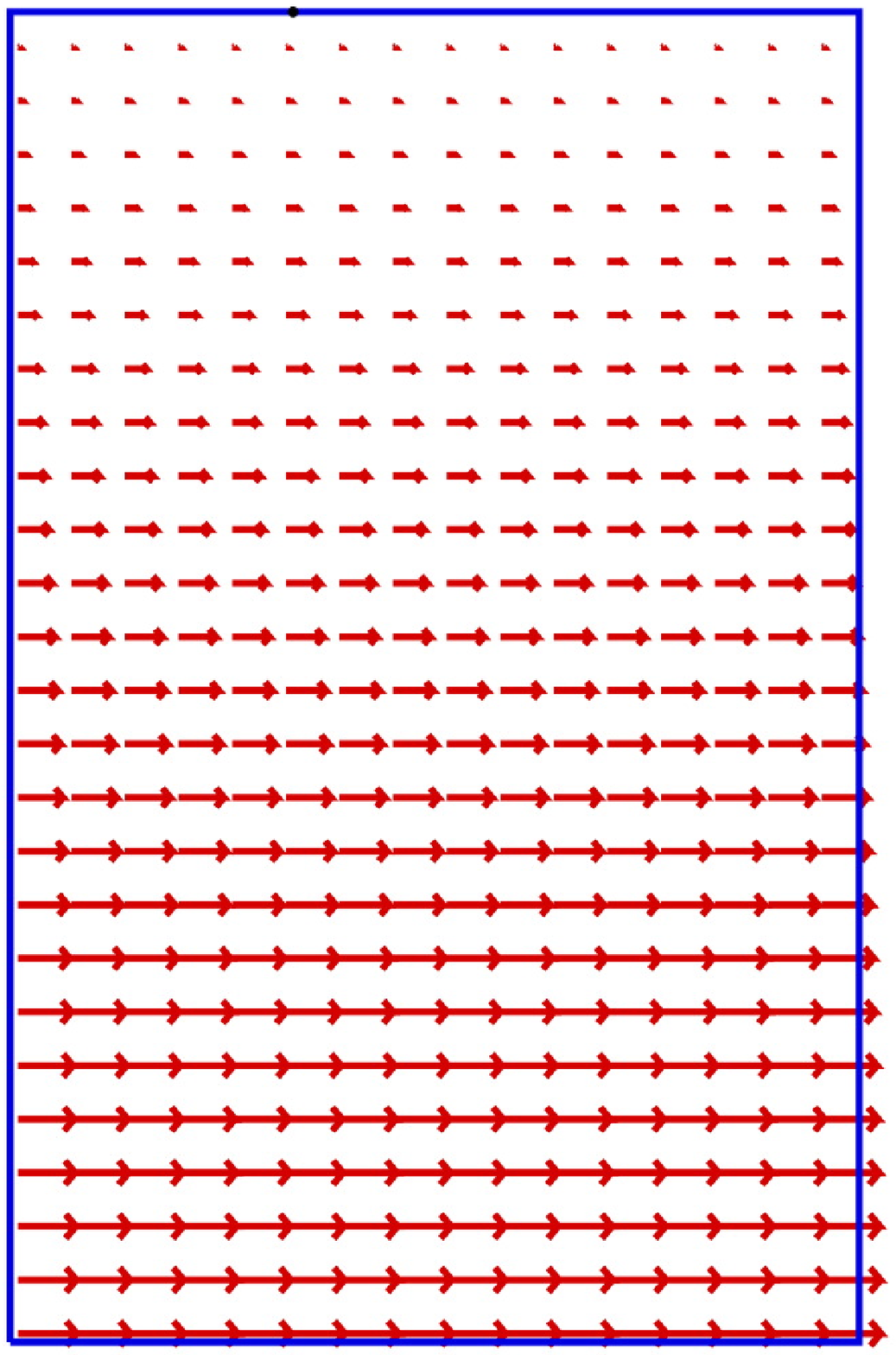}\hspace{2cm}
	\includegraphics[width=4cm]{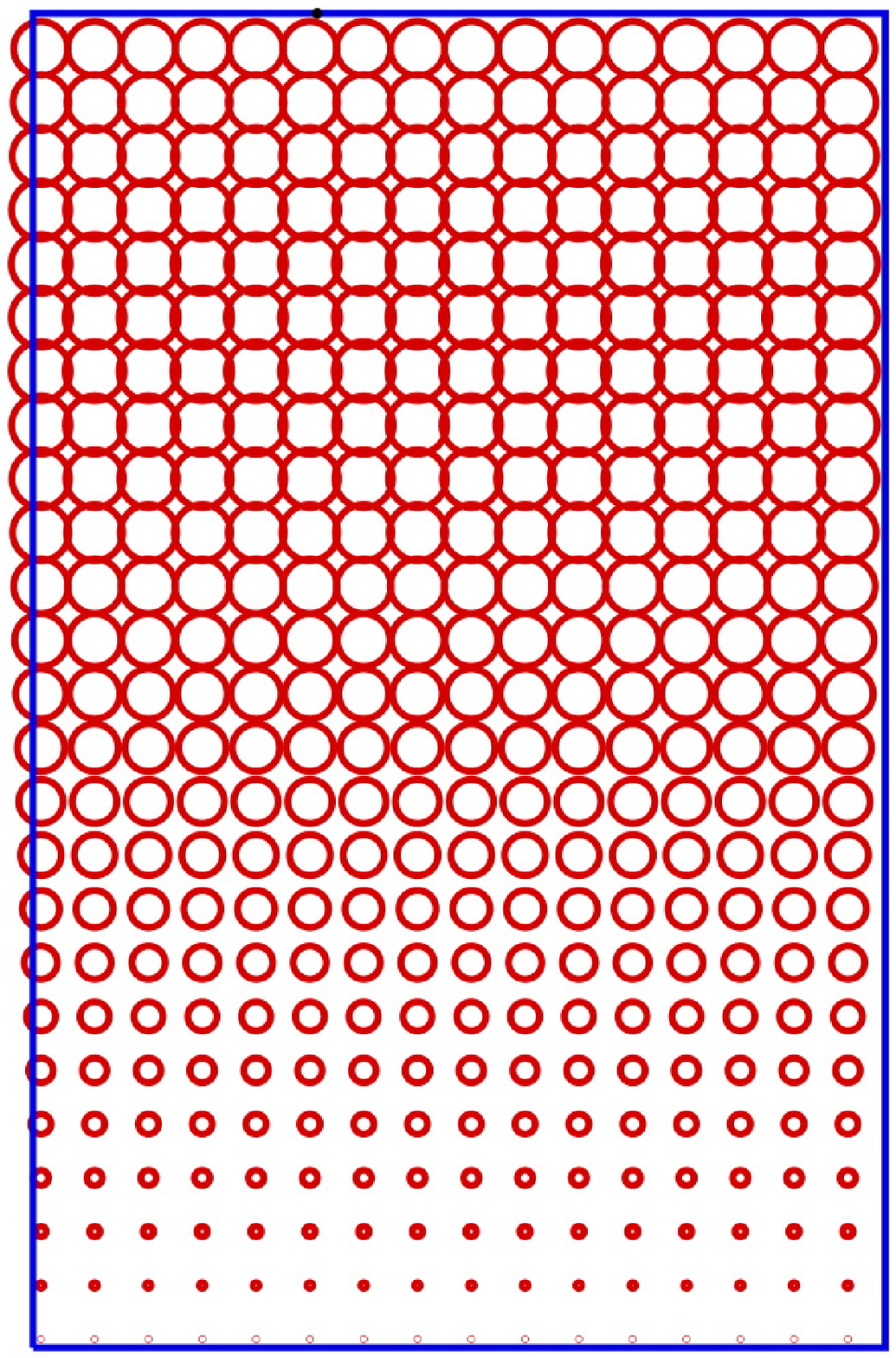}
	\caption{Electric (left) and magnetic (right) fields in a pill-box cavity}
	\label{fig:pillbox}
	\end{center}
	\end{figure}

\noindent The electric and magnetic field pattern of the TM$_{\,010}$ mode can be seen in Fig.~\ref{fig:pillbox}.

\begin{figure}[h!]
	\begin{center}
	\includegraphics[width=2cm]{r-z-coordinates}
	\includegraphics[width=4cm]{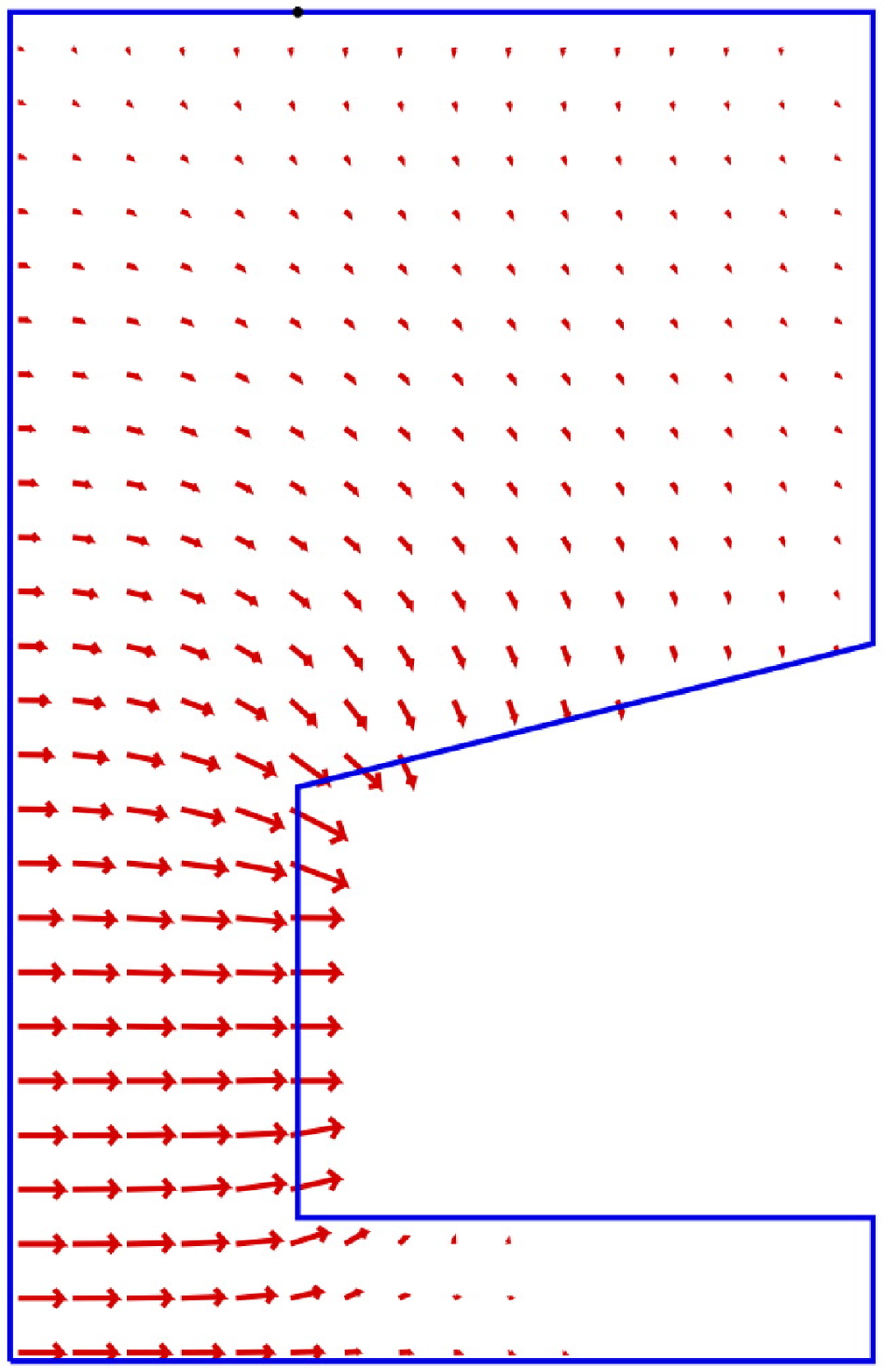}\hspace{2cm}
	\includegraphics[width=4cm]{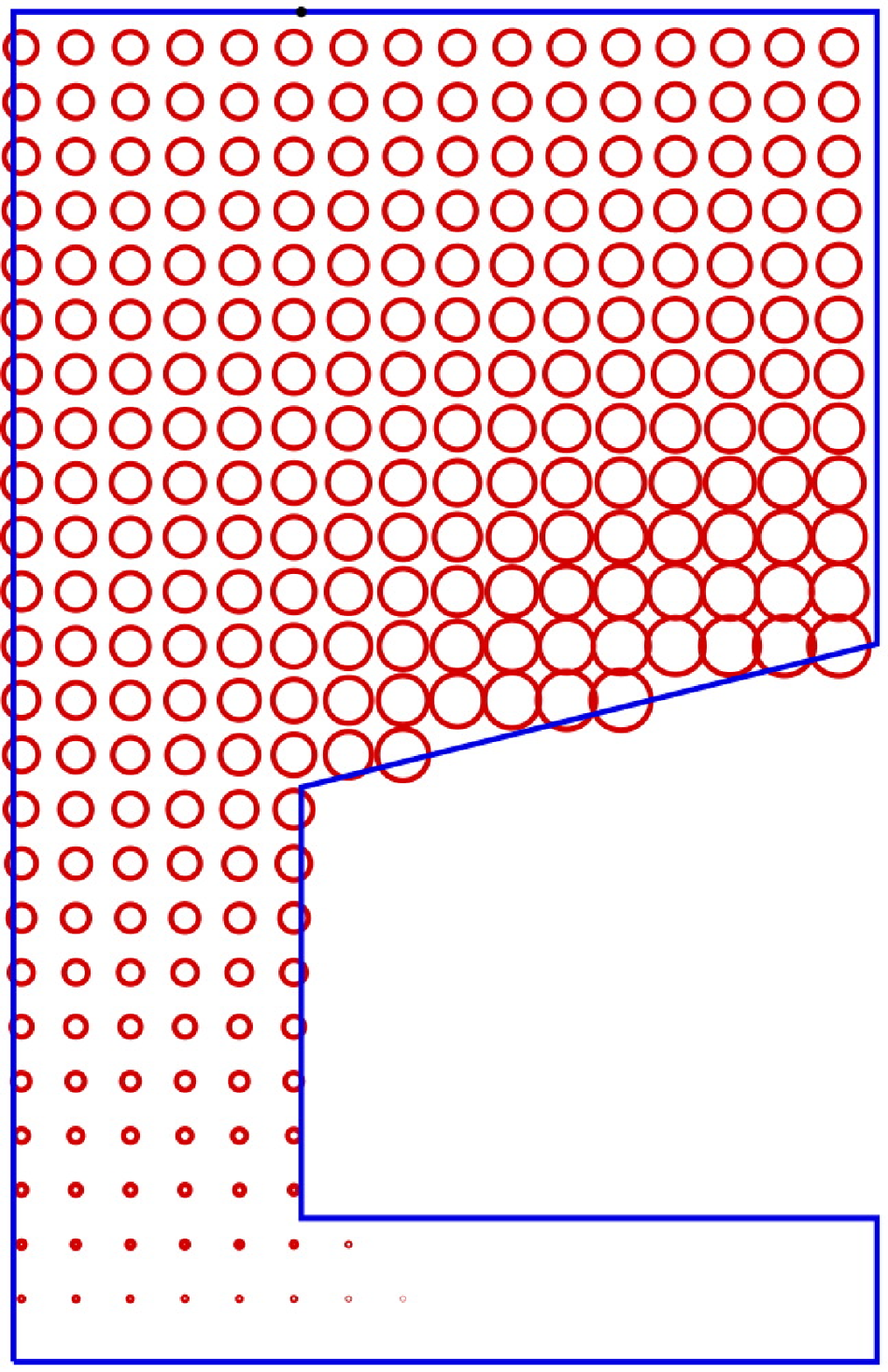}
	\caption{Electric (left) and magnetic (right) fields in a `pill-box-type' cavity with nose cones}
	\label{fig:nosebox}
	\end{center}
	\end{figure}

If one makes a hole on each side of the cavity then particles can travel along the cavity axis in the $z$-direction. However, to increase the acceleration efficiency in normal-conducting standing wave cavities one introduces some kind of `nose cone' on each side of the accelerating gap, which leads to geometries as shown in Fig.~\ref{fig:nosebox}.

These `noses' concentrate the electric field on the beam-axis by increasing the capacitance in this area. At the same time the gap is shortened, which means that the transit time factor increases. For the definition of the transit time factor please consult Ref.~\cite{CASErk}. Another advantage of increasing the capacitance is that according to\\
 	\parbox{9cm}{\[
	\omega_0 = \frac{1}{\sqrt{LC}}\mbox{,}
	\]}\hspace{1.0cm}\parbox{5cm}{\raggedright\bf angular frequency of pill-box cavity}\hfill
	\parbox[b]{8mm}{\begin{equation} \label{eq:fpillboxLC} \end{equation}}

\noindent the inductance has to be lowered to maintain the resonant frequency. Reducing the inductance basically means reducing the volume of the stored magnetic energy and thus having a smaller overall cavity size. To illustrate this statement we consider the magnetic field pattern in Fig.~\ref{fig:nosebox}. The inductance can be calculated via

\[L=\frac{\psi}{I}=\frac{\int \mathbf{B} \cdot d \mathbf{S}}{\oint \mathbf{H} \cdot d\mathbf{l}}\mbox{.} \] 

The magnetic flux $\psi$ is calculated via the surface integral of the magnetic field which penetrates the cross-section of the cavity as shown in Fig.~\ref{fig:nosebox}. The electric current is flowing on the inner surface of the cavity, thus we can calculate it via a closed line integral around the cavity shape of Fig.~\ref{fig:nosebox}. This means that reducing the cross-section of the cavity will reduce the surface which is penetrated by magnetic field lines and thus we obtain a smaller inductance. One can also say that the inductance goes down because the amount of stored magnetic energy is reduced for a smaller cavity radius. 

Similarly one can calculate the capacitance via

\[C= \frac{Q}{V} = \frac{\oint \mathbf{D} \cdot d\mathbf{S}}{\int \mathbf{E}\cdot d\mathbf{l}}\mbox{.}\]

In the case of a classic plate capacitance the stored charge $Q$ is given by the surface integral enclosing one of the capacitor plates with the normal vector of the surface pointing in the same direction as the electric field lines ending (or starting) on the plate. In the case of the cavity in Fig.~\ref{fig:nosebox} we have to integrate along the surface of one half-cavity. The voltage $V$ is then simply the line integral along the electric field lines between the noses of the cavity. 

Another method to determine the capacitance and inductance of a cavity is the following: for an arbitrary cavity shape that one has simulated with any 2D or 3D simulation code one can determine $C$ and $L$ of a particular cavity mode from its $(R/Q)$ value and the frequency by using

\[\omega= 2\pi f = \frac{1}{\sqrt{LC}} \hspace{1cm} \mbox{and} \hspace{1cm} \left(\frac{R}{Q}\right) = \sqrt{\frac{L}{C}}\mbox{.}\]

Even though we have changed the field pattern in the cavity (the electric field is no longer parallel to the axis in all parts of the cavity) we still have a TM$_{\,010}$ mode in the cavity (see definition of TM below). 

\subsection{Nomenclature of electromagnetic modes}
\label{sec:Nomenclature}
The definition of modes is generally done according to the following rules:

\noindent{\bf TM{\boldmath$_{mnp}$} or E{\boldmath$_{mnp}$} modes}
\begin{itemize}
\item magnetic field components only in transverse direction (TM --- transverse magnetic),
\begin{itemize}
\item[$\rightarrow$] no longitudinal magnetic field component ($B_z=0$),
\item[$\rightarrow$] non-vanishing longitudinal electric field component ($E_z\neq 0$);
\end{itemize}
\end{itemize}

\noindent{\bf TE{\boldmath$_{mnp}$} or H{\boldmath$_{mnp}$} modes}
\begin{itemize}
\item electric field components only in transverse direction (TE --- transverse electric),
\begin{itemize}
\item[$\rightarrow$] no longitudinal electric field component ($E_z=0$),
\item[$\rightarrow$] non-vanishing longitudinal magnetic field component ($B_z\neq 0$).
\end{itemize}
\end{itemize}

\noindent In cylindrical coordinates the indices $m$, $n$, $p$ are:
\begin{itemize}
\item[$m$] number of full-period variations of the field components in the azimuthal direction, in cylindrical resonators this means: $\mathbf E$, $\mathbf B \propto$ $\cos(m\varphi)$ or $\sin(m\varphi)$;
\item[$n$] number of zero-crossings of the longitudinal field components in the radial direction, in cylindrical resonators this means: $E_z$, $B_z \propto$ $J_m(x_{mn} r/R_c) $, $x_{mn}$ are the zeros of the $J_m$;
\item[$p$] number of half-period variations of the field components in the longitudinal direction, in cylindrical resonators this means: $\mathbf E$, $\mathbf B \propto$ $\cos(p\pi z/l)$ or $\sin(p\pi z/l)$.
\end{itemize}

\begin{figure}[h!]
	\begin{center}
	\includegraphics[width=2cm]{r-z-coordinates}
	\includegraphics[height=4.5cm]{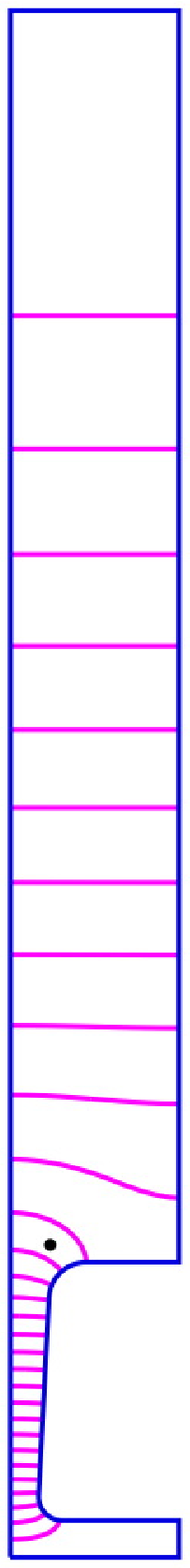}\hspace{1cm}
	\includegraphics[height=4.5cm]{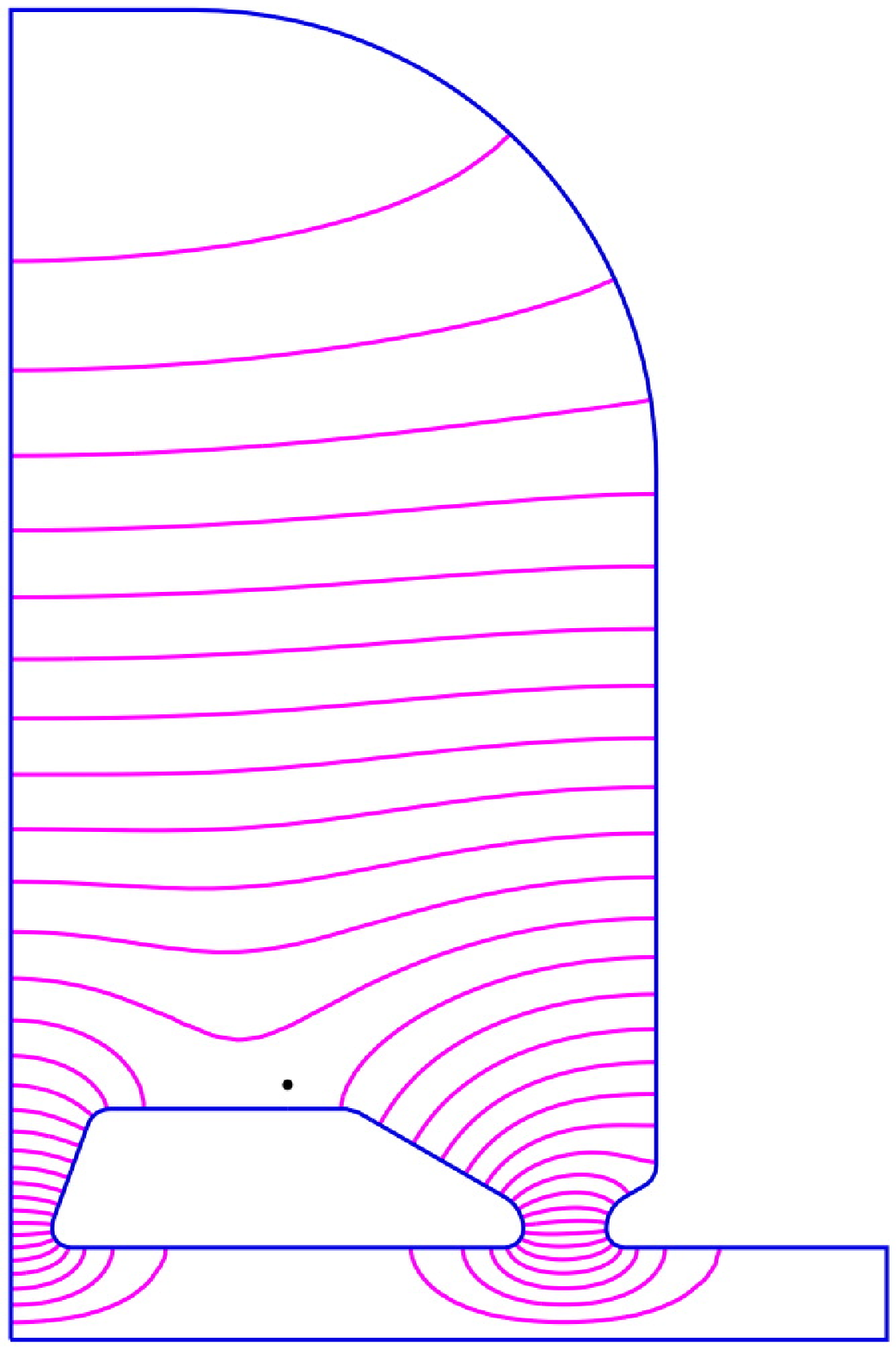}\hspace{1cm}
	\includegraphics[height=4.5cm]{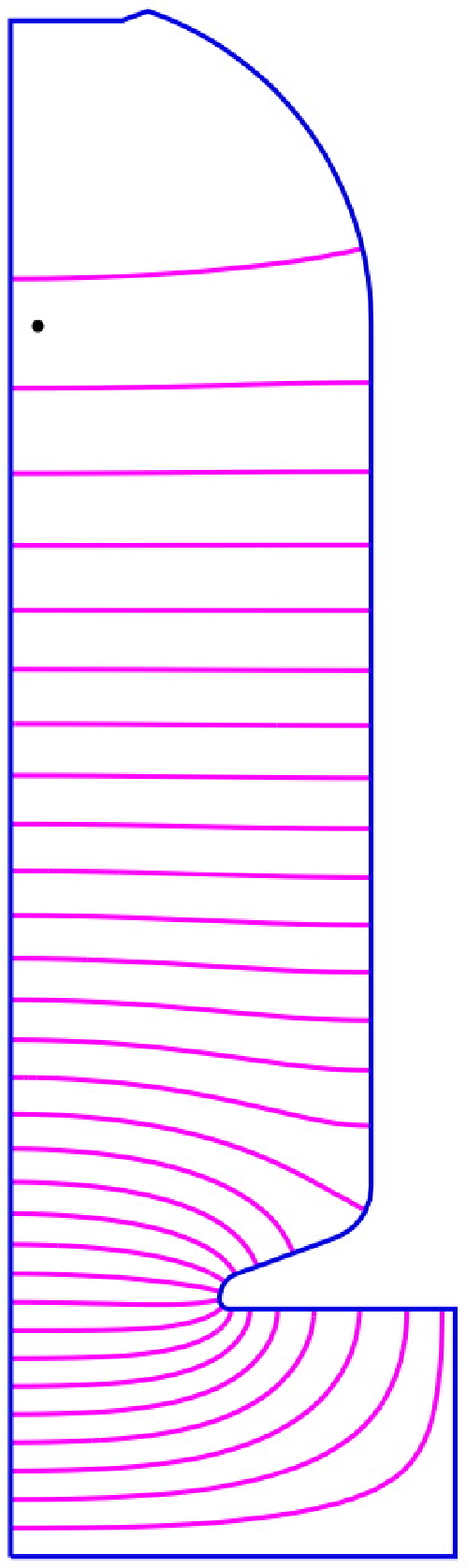}\hspace{1cm}
	\includegraphics[height=4.5cm]{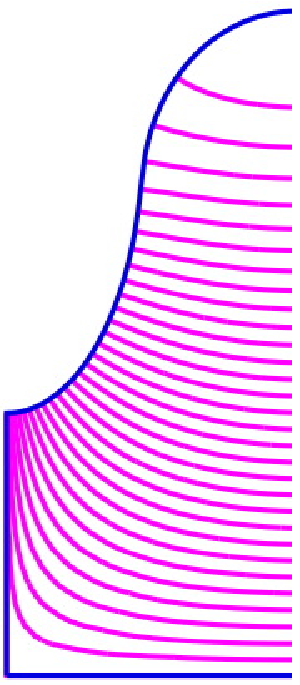}
	\caption{Typical TM mode cavities, from left to right: Drift Tube Linac (DTL), Cell-Coupled Drift Tube Linac (CCDTL), Cell-Coupled Linac (CCL), superconducting elliptical cavities}
	\label{fig:TMtypical}
	\end{center}
	\end{figure}

\noindent Some typical examples of TM mode cavities are shown in Fig.~\ref{fig:TMtypical}.

\subsection{TE mode cavities}
\begin{figure}[h!]
	\begin{center}
	\includegraphics*[width=\textwidth]{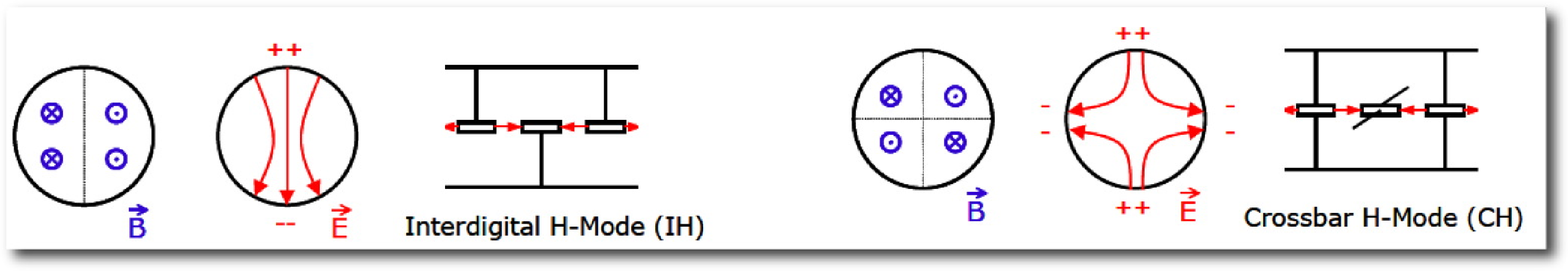}
	\caption{Commonly used TE mode cavities, left: TE$_{110}$ (interdigital TE mode), right: TE$_{210}$ (crossbar TE mode) \cite{Tiede08}}
	\label{fig:TE-modes}
	\end{center}
	\end{figure}

TE or H mode cavities are an interesting species, because by definition TE modes do not have a longitudinal electrical field component, which could be used for acceleration. On the other hand, these modes have much lower surface losses than TM modes because there is less magnetic field on the circumference of the cavity (close to the conducting walls). Less magnetic field means less surface current and hence lower losses on the inner surface of the cavity. To use the advantage of lower losses the electric field lines of the TE modes are `bent' in such a way that we get an axial electric field component, which can be used for acceleration. This is usually done by introducing drift tubes as shown in Fig.~\ref{fig:TE-modes}.

\begin{figure}[h!]
	\begin{center}
	\includegraphics[width=0.6\textwidth]{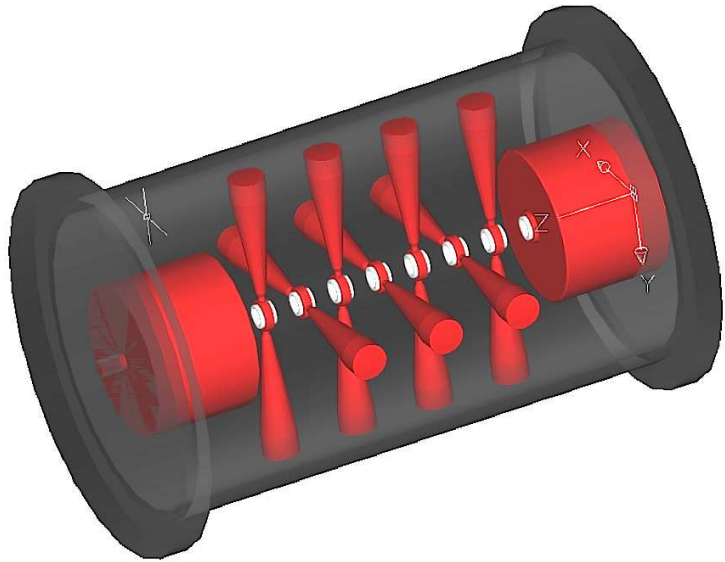}
	\caption{Crossbar H-mode cavity design example \cite{ClementeThesis}}
	\label{fig:H-mode-cavity}
	\end{center}
	\end{figure}

\begin{figure}[h!]
	\begin{center}
	\includegraphics*[width=0.85\textwidth]{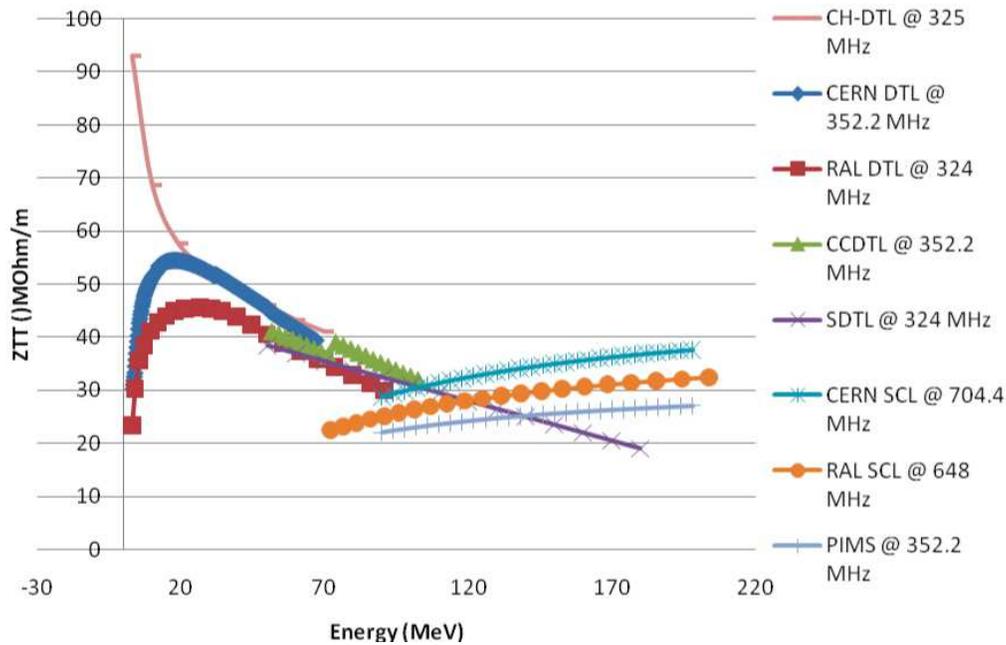}
	\caption{Shunt impedance comparison for various normal-conducting structures \cite{Ciprian08}}
	\label{fig:TEshunt}
	\end{center}
	\end{figure}

Strictly speaking these modes are no longer pure TE modes because we now have a longitudinal electric field on axis. The dominant field distribution, however, remains that of a TE mode, which means that one can take advantage of the low losses on the inner surface. The design of TE mode cavities is further complicated by the fact that a transverse electric field cannot exist parallel to the end walls of a conducting cavity. By definition electric field lines can only have a normal orientation with respect to conducting surfaces, which means that the end-cells of TE mode cavities need a special design effort to allow for the existence of a TE mode within the cavity (see Fig.~\ref{fig:H-mode-cavity}). 
As a consequence TE mode cavities can only make use of their lower surface losses if they consist of many cells per cavity.  If we compare a typical shunt impedance curve for TE mode structures with traditional TM mode cavities (see Fig.~\ref{fig:TEshunt}), we can see that 
the advantage of lower shunt impedance only persists at low energy (for protons approximately below 20\,MeV--30\,MeV, \cite{Ciprian08}). Since at low energy space-charge forces usually enforce short transverse focusing periods, it is difficult to have a large number of accelerating cells per cavity. This problem was solved rather elegantly by the invention of the KONUS (KOmbinierte NUll grad Struktur, \cite{RatzingerHabil}) beam dynamics principle, which reduces the transverse RF defocusing and which made it possible to have longer transverse focusing periods. This principle was applied successfully to several low-duty-cycle heavy-ion accelerators, for instance Linac3 \cite{Linac3} and REX-ISOLDE \cite{Podlech99} at CERN. Up to now TE mode cavities in combination with the KONUS beam dynamics have not been used in high (average) beam power applications, where beam loss is a potentially performance limiting issue. At present several design proposals have been submitted for high-beam-power applications. 

\subsection{TEM mode cavities}
\begin{figure}[h!]
	\begin{center}
	\includegraphics[width=0.8\textwidth]{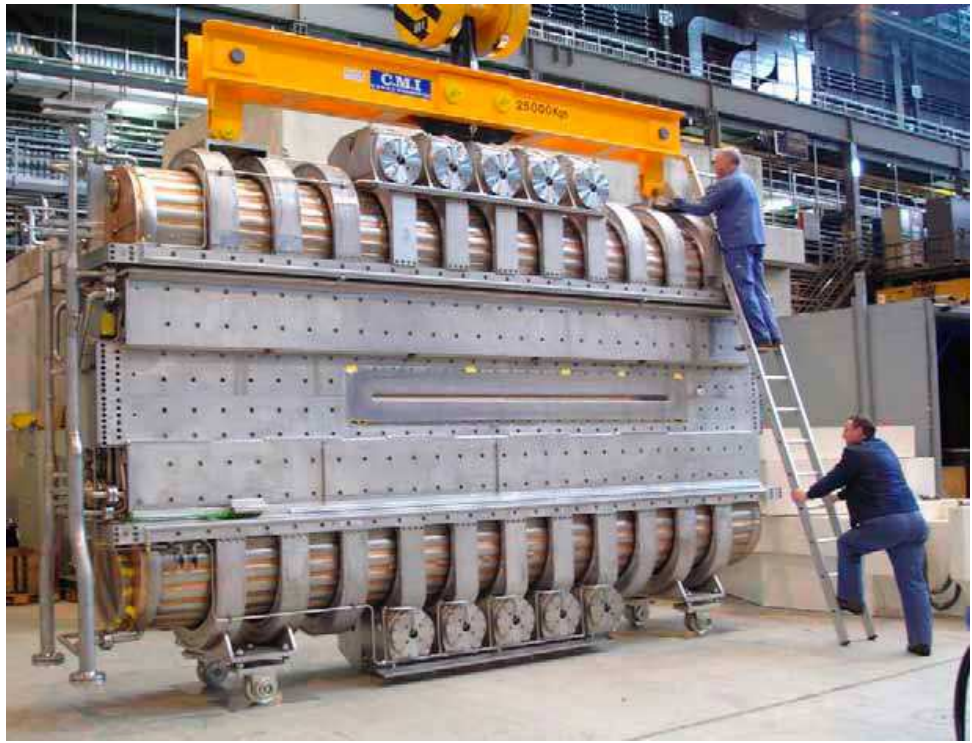}
	\caption{PSI cyclotron TM-mode cavity working at 50.6\,MHz \cite{HFitze}}
	\label{fig:PSI-cavity}
	\end{center}
	\end{figure}

The frequency of the accelerating modes used in TE and TM mode cavities is related to the transverse dimensions of these cavities and the capacitance between the drift tubes or nose cones. As a consequence, practically usable dimensions are related to a certain frequency range, which typically starts at around 100\,MHz and extends up to 10s of GHz for travelling wave structures. Lower frequencies yield prohibitively large cavities, and higher frequencies impose unrealistically tight tolerances on beam steering, alignment, and mechanical construction. An example of a very low frequency cavity is shown in Fig.~\ref{fig:PSI-cavity}, which depicts a 50\,MHz cyclotron cavity used at PSI in Switzerland \cite{HFitze}. The cavity is 5.6\,m long, provides an accelerating voltage of 1\,MV and weighs 25\,tons.

\begin{figure}[h!]
	\begin{center}
	\includegraphics[width=0.35\textwidth]{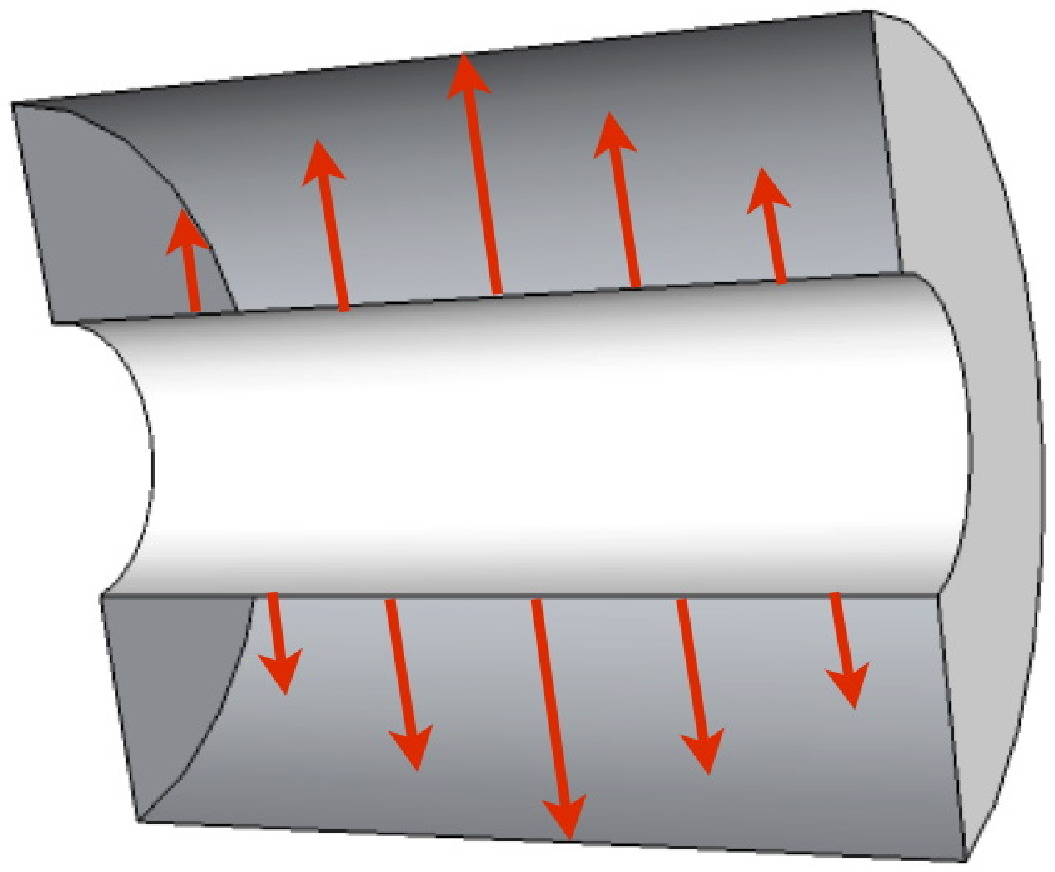}
	\includegraphics[width=0.48\textwidth]{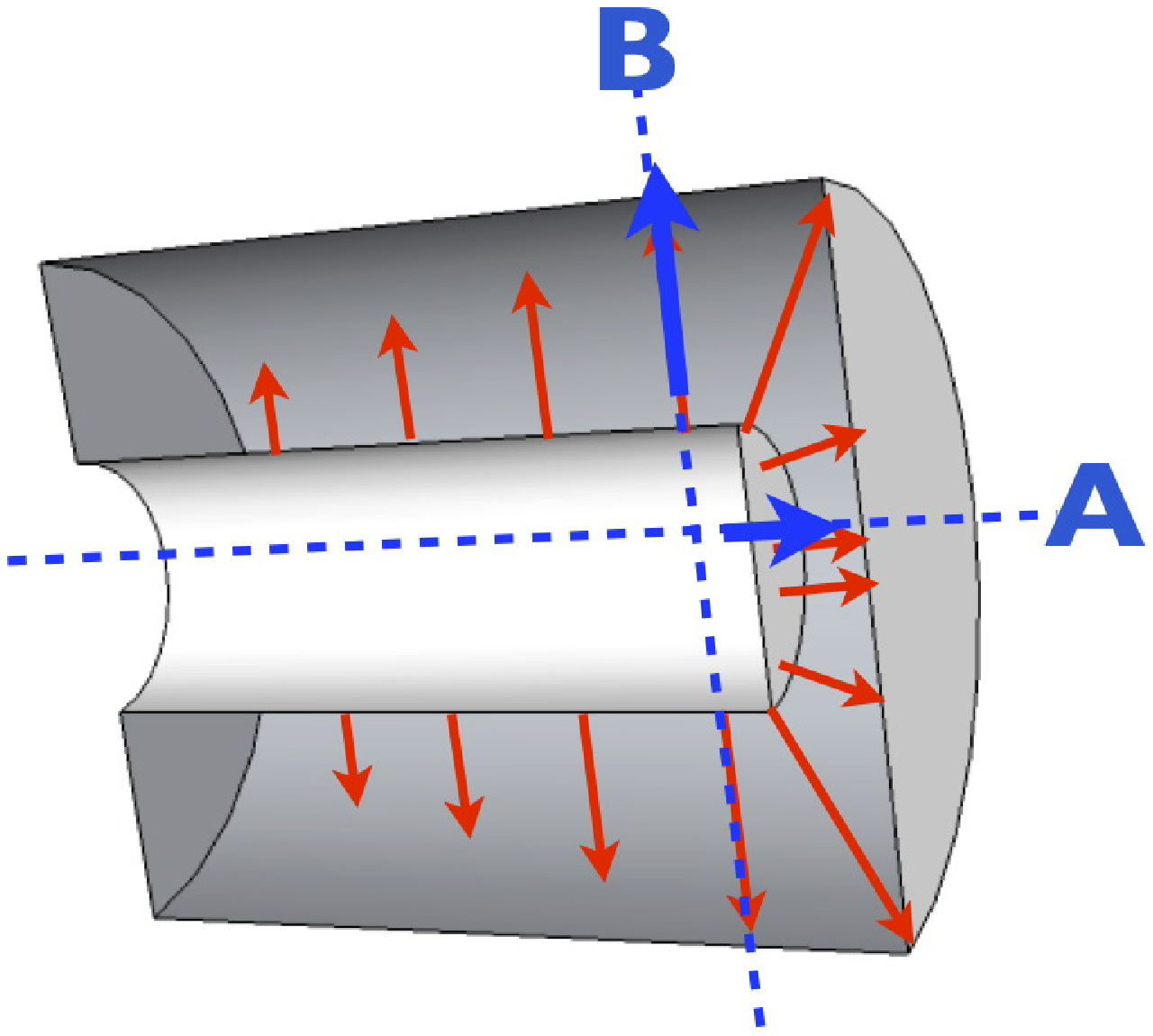}
	\caption{Left: coaxial 1/2 wave cavity, right: 1/4 wave cavity}
	\label{fig:TEM-mode}
	\end{center}
	\end{figure}

For certain applications, such as small- to medium-sized synchrotrons, one needs frequencies in the MHz range, and it is clear that TM or TE-mode cavities would become excessively large. Here one can make use of TEM cavities, were the RF frequency is no longer determined by the transverse dimensions of the cavity but rather by its length. A typical example is a coaxial cavity as shown in Fig.~\ref{fig:TEM-mode}, where the length of the cavity equals one half of a wave-length in longitudinal direction. 

Instead of using an electric wall as boundary condition on both sides of the cavity one can also use an `open' boundary on one end of the cavity. This has the benefits of i) shortening the length of the cavity to a 1/4 wave-length, and of ii) bending the electric field lines into the direction of the cavity axis so that one can accelerate particles along the cavity axis (direction A in Fig.~\ref{fig:TEM-mode}). In addition to direction A one can also accelerate in direction B, which requires a tight synchronization between the particles and the RF. Direction A is typically used in normal-conducting synchrotron cavities, while direction B is often used in normal and superconducting Quarter-Wave and Half-Wave Resonators (QWR and HWR) in low-frequency ion linacs. Quarter-wave cavities are of course no longer real TEM cavities, but since they originate from a TEM mode the classification seems somewhat justified. 

\begin{figure}[h!]
	\begin{center}
	\includegraphics[width=0.27\textwidth]{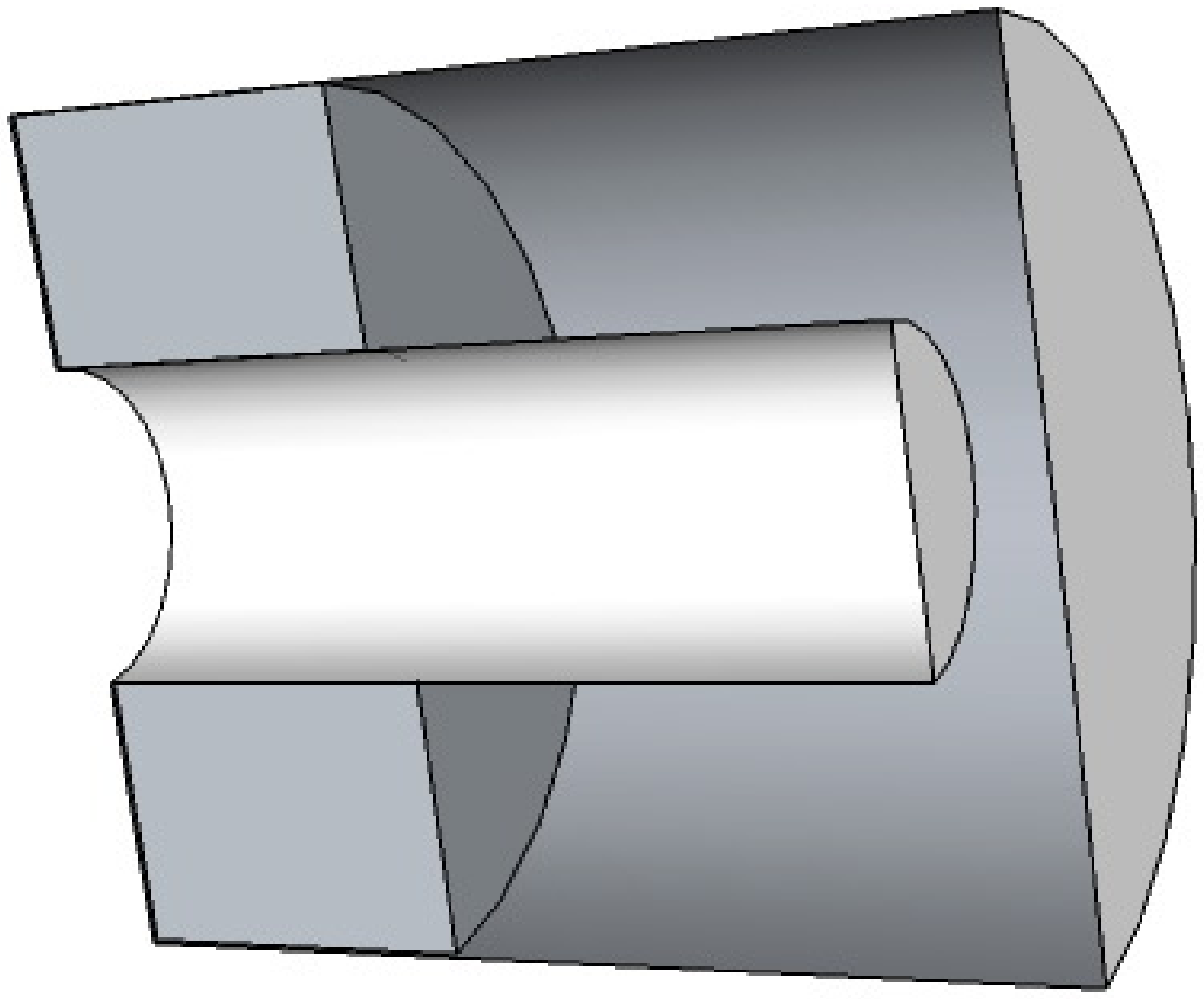}\hspace{2cm}
	\includegraphics[width=0.4\textwidth]{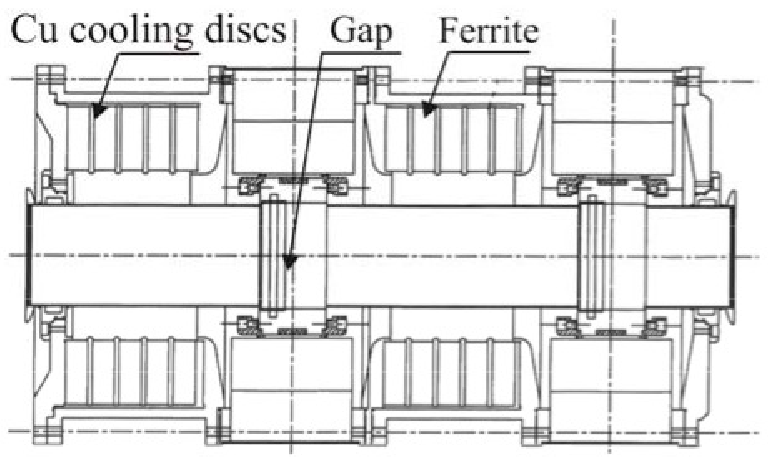}
	\caption{Left: principle of `filled' Quarter-Wave Resonator, right: actual cross-section of a 13-20\,MHz cavity in the CERN PS \cite{Morvillo}}
	\label{fig:filled-coax}
	\end{center}
	\end{figure}

In the case of synchrotron cavities one often fills part of the volume between inner and outer conductor with a dielectric or magnetic material as shown in Fig.~\ref{fig:filled-coax}. Both material types will shorten the cavity because of their dielectric/magnetic material constants. Magnetic materials like ferrites have the added advantage that the frequency of the cavity can be changed by the application of external fields. Because of the losses in the material the quality factor $Q$ becomes very low, which means that only a small amount of energy is stored in the cavity enabling a fast frequency change. These cavities can thus be used in low-energy synchrotrons as outlined in Section~\ref{sec:velvar}. 

\begin{figure}[h!]
	\begin{center}
	\includegraphics[width=0.6\textwidth]{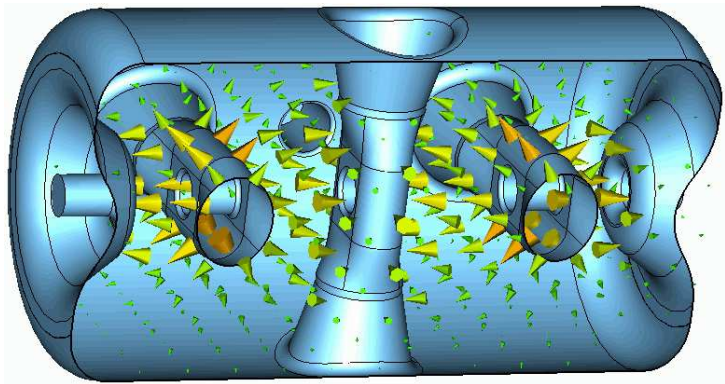}
	\caption{Triple-spoke cavity developed at FZJ \cite{Zaplatin}}
	\label{fig:spoke}
	\end{center}
	\end{figure}

An interesting variant of a TEM cavity, which is in fact difficult to identify as such, is called a `spoke cavity' \cite{Delayen88}. It consists of $1\ldots n$ stacked half-wave cavities, with the inner conductor (the spoke) being rotated by $90^{\circ}$ from cell to cell. Figure~\ref{fig:spoke} shows an example of a triple spoke cavity developed at Forschungszentrum J\"{u}lich \cite{Zaplatin}.
At first sight spoke cavities are very similar to crossbar H-mode cavities as in Fig.~\ref{fig:H-mode-cavity}, and even the field distribution does not show striking differences. Nevertheless, the modes in both cavities are quite different: in the H-mode cavities the `bars' are quite thin in order not to disturb too much the H-mode, which is determined mainly by the diameter of the cavity. In spoke cavities the field distribution is determined mainly by the spokes, which represent sections of short-circuited coaxial resonators, hence they are very `thick' in comparison to the H-mode cavities. Another difference is the presence of the modified end-cells in the H-mode cavities, which are needed to make the existence of H-modes in cylindrical cavities possible. The `nose-cones', in the spoke cavity end-walls are introduced to optimize the transit time factor, but they are not needed to establish the desired field distribution. Even though many prototypes of spoke cavities have already been built, they have not yet been used in an actual accelerator. Several projects, however, have foreseen them in the low-energy parts of ion and proton accelerators, and we should soon see the first spoke cavities in operation. 

\section{Normal versus superconducting cavities}
When considering normal or superconducting cavities for a particular application one should keep in mind that superconducting does not automatically mean better or more efficient than normal-conducting. There are many instances where normal-conducting cavities can not only be more efficient but can also reach higher gradients than superconducting cavities. The purpose of this section is to highlight the different design approaches for both cavity types and to give some guidelines on efficiency as a function of the operational parameters. 

\subsection{Design of normal and superconducting cavities}
\begin{figure}[h!]
	\begin{center}
	\includegraphics[width=2cm]{r-z-coordinates}
	\includegraphics[height=4cm]{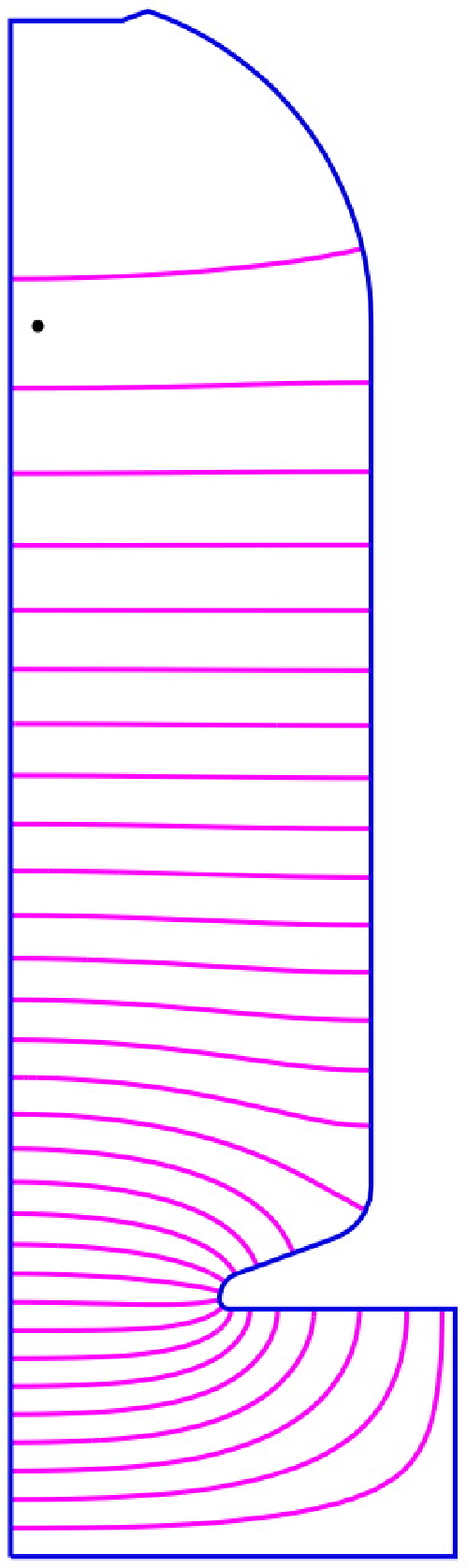}\hspace{2.5cm}
	\includegraphics[height=4cm]{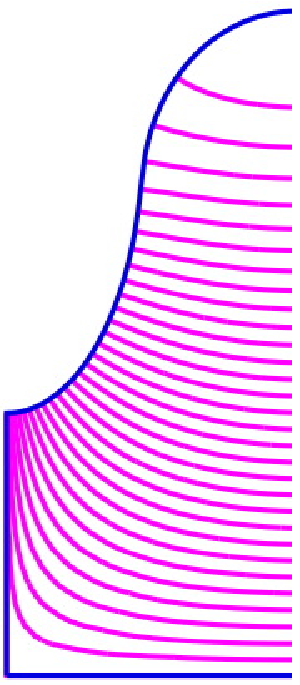}
	\caption{Left: normal-conducting cavity with nose cones, right: superconducting elliptical cavity}
	\label{fig:normalvssuper}
	\end{center}
	\end{figure}

In Fig.~\ref{fig:normalvssuper} we can see two cavity shapes, which can be regarded as representative of normal and superconducting cavities. From the different shapes one can see that the criteria to optimize the cavity geometry must have been quite different and in this section we will see why that is so.

The efficiency of any cavity is given by the ratio of the (average) power, which is is used to accelerate the beam and the (average) power needed from the RF generator:\\
 	\parbox{9cm}{\[
	\eta = \frac{\overline{P_{beam}}}{\overline{P_{gen}}} = \frac{\overline{P_{beam}}}{\overline{P_{beam}}+\overline{P_{d}}}\mbox{.}
	\]}\hspace{1.0cm}\parbox{5cm}{\raggedright\bf efficiency of cavities}\hfill
	\parbox[b]{8mm}{\begin{equation} \label{eq:avefficiency} \end{equation}}
	
To calculate the average power which is lost in the cavity walls we first need the instantaneous power which is dissipated on the cavity surface. In the case of normal-conducting cavities one usually defines\\
 	\parbox{9cm}{\[
	P_d = \frac{V_{acc}^2}{ZT^2 L}\mbox{.}
	\]}\hspace{1.0cm}\parbox{5cm}{\raggedright\bf power dissipated in normal-conducting cavities}\hfill
	\parbox[b]{8mm}{\begin{equation} \label{eq:pdnormal} \end{equation}}

From this definition it is clear that a high effective shunt impedance $ZT^{\,2}$ will minimize the power loss and thus normal-conducting cavities usually have nose cones, which increase the transit time factor and which increase the peak fields around the axis. In the design process one generally increases the peak fields in the nose area up to the highest viable value, in order to obtain a high shunt impedance. 

The limits of peak fields are qualified according to a definition from 1957, where Kilpatrick \cite{Kilpatrick} defined the maximum achievable field in vacuum as a function of frequency as\\
 	\parbox{9cm}{\[
	f \mbox{ [MHz]}=1.64 E_k^2 e^{-8.5/E_k} \mbox{, $E_k$ in [MV/m]}\mbox{.}
	\]}\hspace{1.0cm}\parbox{5cm}{\raggedright\bf Kilpatrick limit}\hfill
	\parbox[b]{8mm}{\begin{equation} \label{eq:kilpatrick} \end{equation}}

In modern cavities, however, one surpasses this limit by a certain `bravery factor' $b$:  $E_{\,surf} = b \cdot E_{k}$, which is usually between 1 and 2 (one speaks of units of `Kilpatrick') for most normal-conducting cavities. Surpassing the Kilpatrick limit became possible by having better vacuum and cleaner surfaces on the inside of the cavities. For frequencies above 1\,GHz and for very short pulses the Kilpatrick limit has basically lost its importance because the maximum achievable fields are governed by different phenomena. 

In superconducting cavities the situation is different. The instantaneous power delivered to the beam is for most applications much higher than the surface losses ($P_{beam} \ll P_{d}$). Since the surface losses are so small it becomes possible to use cavities with very high accelerating gradients, since the RF system is  no longer limited by losses, which rise quadratically with the cavity voltage [see Eq.~\eqref{eq:pdnormal}]. So instead of optimizing for high shunt impedance one now optimizes the cavity shape for low ratios of $E_{peak,\,surf}/E_{acc}$ and $B_{peak,\,surf}/E_{acc}$, which explains the elliptical cavity shape of basically all high-gradient superconducting cavities, which are used for highly relativistic particles. Nose cones would increase the peak fields, so instead one tries to minimize the peak fields on the inner surface. In the case of elliptical cavities for particles close to the speed of light ($\beta = 1$), one can achieve a minimum ratio of peak surface field to accelerating field of $E_{peak,\,surf}/E_{acc} \approx 2$. 

Apart from the RF power, one also has to consider the cryogenic power needed to cool the surface losses during the presence of RF fields in the cavity.  In superconducting cavities the surface losses are usually expressed as \\ 
 	\parbox{9cm}{\[
	P_d = \frac{V_{acc}^2}{(R/Q)Q_0}\mbox{.}
	\]}\hspace{1.0cm}\parbox{5cm}{\raggedright\bf power dissipated in superconducting cavities}\hfill
	\parbox[b]{8mm}{\begin{equation} \label{eq:pdsuper} \end{equation}}

In order to reduce the cryogenic power one can change the quality factor $Q_{\,0}$ of the cavity by optimizing the cryogenic temperature for the required operational parameters.

\subsection{Superconducting cavities in pulsed operation}
At first sight superconducting cavities seem extremely attractive because their high $Q_{\,0}$ values in the range of $10^9$ to $10^{10}$ (instead of $\approx 10^4$ for normal-conducting cavities) basically reduce the surface losses down to almost zero. Having a high $Q_0$, however, not only means having low surface losses but it also implies a high amount of stored energy\\
 	\parbox{9cm}{\[
	Q_0= \frac{\omega_0 W}{P_d}
	\]}\hspace{1.0cm}\parbox{5cm}{\raggedright\bf unloaded Q}\hfill
	\parbox[b]{8mm}{\begin{equation}  \end{equation}}
which has to be put into the cavity before the arrival of the beam. In other words superconducting cavities often have a long `filling time', which means that the RF pulse length has to be considerably longer than the actual beam pulse length. As a consequence the efficiency of the RF system is reduced! In the following you will find some formulae that allow a quick calculation of the extra power which is needed to operate superconducting cavities in pulsed mode (taken from Ref.~\cite{Gerigk06}). 

The filling time constant can be calculated as\\
  	\parbox{9cm}{\[
	\tau_l = \frac{Q_l}{\omega_0}=\frac{Q_0}{\omega_0(1+\beta)} \approx \frac{Q_0}{\omega_0 \cdot P_{beam}/P_d}\mbox{,}
	\]}\hspace{1.0cm}\parbox{5cm}{\raggedright\bf filling time constant in superconducting cavities}\hfill
	\parbox[b]{8mm}{\begin{equation}  \end{equation}}

\noindent where $Q_l$ is the quality factor of the loaded cavity (`loaded' with the beam and coupled to a wave-guide) and $\beta$ is the coupling factor of the power coupler (not to be confused with the relativistic $\beta$, which defines the velocity of the beam). Please note that some authors define the filling time as $\tau_{\,l}=2Q_l/\omega_0$. In the above formula we have used the following simplification that can only be used for superconducting cavities:\\
  	\parbox{9cm}{\[
	\beta = 1+ \frac{P_{beam}}{P_d} \approx \frac{P_{beam}}{P_d}\mbox{.}
	\]}\hspace{1.0cm}\parbox{5cm}{\raggedright\bf definition of the coupling factor in superconducting cavities}\hfill
	\parbox[b]{8mm}{\begin{equation}  \end{equation}}

\noindent Using\\
  	\parbox{9cm}{\[
	P_{beam} = I_{beam}V_{acc}\cos{\phi_s}
	\]}\hspace{1.0cm}\parbox{5cm}{\raggedright\bf power given to the beam}\hfill
	\parbox[b]{8mm}{\begin{equation}  \end{equation}}

\noindent one can calculate the filling time of superconducting cavities from\\
  	\parbox{9cm}{\[
	\tau_l \approx \frac{V_{acc}}{\omega_0(R/Q) I_{beam}\cos{\phi_s}}\mbox{,}
	\]}\hspace{1.0cm}\parbox{5cm}{\raggedright\bf filling time constant in superconducting cavities}\hfill
	\parbox[b]{8mm}{\begin{equation}  \end{equation}}
	
\noindent which allows us to calculate the cavity voltage profile during the filling process.\\

	\vspace{-1cm}
  	\parbox{9cm}{\[
	V_{c,\,filling}(t) = 2 V_0 \left( 1-e^{-\frac{t}{2\tau_l}} \right)\mbox{.}
	\]}\parbox{5cm}{\raggedright\bf cavity voltage during filling}\hfill
	\parbox[b]{8mm}{\begin{equation}  \label{eq:cavdfil} \end{equation}}
	
Equation \eqref{eq:cavdfil} also allows us to calculate the total filling time for a superconducting cavity. To do this we assume that the generator delivers exactly the voltage needed to establish the desired cavity voltage as\\  

	\vspace{-1cm}
  	\parbox{9cm}{\[
	t=\tau_l \ln (4)\mbox{.}
	\]}\parbox{5cm}{\raggedright\bf total filling time of a superconducting cavity}\hfill
	\parbox[b]{8mm}{\begin{equation} \label{eq:totalfill} \end{equation}}

In a real set-up the RF source will usually have some power margin allowing for a faster filling time (but still needing the same amount of energy to fill the cavity). The reflected voltage during the filling process is determined by $V_c (t) = V_0 + V_r (t)$, where $V_0$ stands for the nominal cavity voltage. Thus we can write\\

	\vspace{-1cm}
  	\parbox{9cm}{\[
	V_{r,filling} = V_0 \left(1-2 e^{-\frac{t}{2\tau_l}}\right)\mbox{.}
	\]}\parbox{5cm}{\raggedright\bf reflected voltage during filling}\hfill
	\parbox[b]{8mm}{\begin{equation} \label{eq:vreflfill} \end{equation}}

During the passage of the beam the voltage remains constant at the nominal value until it is switched off immediately after the last bunch has left the cavity. From that point onwards the cavity voltage decays according to\\

	\vspace{-1cm}
  	\parbox{9cm}{\[
	V_{c,\,decay}(t) = V_0 \cdot e^{-\frac{t}{2\tau_l}}\mbox{.}
	\]}\parbox{5cm}{\raggedright\bf cavity voltage decay}\hfill
	\parbox[b]{8mm}{\begin{equation} \label{eq:vdecay} \end{equation}}

\begin{figure}[h!]
	\begin{center}
	\includegraphics[width=0.7\textwidth]{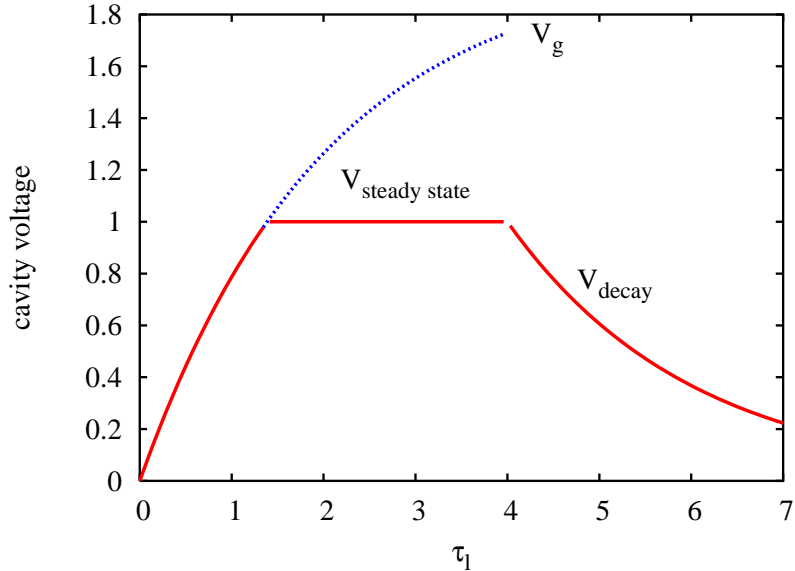}
	\caption{Voltage profile in a superconducting cavity for one beam pulse}
	\label{fig:voltageprofile}
	\end{center}
	\end{figure}

Using Eqs.~\eqref{eq:vreflfill} and \eqref{eq:vdecay} we can plot the voltage profile in a superconducting cavity for one beam pulse as shown in Fig.~\ref{fig:voltageprofile}.

With the above formulas we can now calculate the `wasted' RF energy, which is needed for each pulse to fill the cavity, and also the total cryogenic energy, which is needed to cover the surface losses for filling, beam pulse, and decay of the fields in the cavity. 

We start with the `wasted' RF energy, for which we use the total filling time (\ref{eq:totalfill}) to integrate over the reflected power [square of Eq.~\eqref{eq:vreflfill}] during filling of the cavity:\\
  	\parbox{9cm}{\begin{align*}
	W_{r, \,filling} &= P_{gen} \int^{\ln(4)\tau_l}_0 \left(1-2e^{-\frac{t}{2\tau_l}} \right)^2 dt \\ 
	                    &= P_{gen} \cdot \tau_l \underbrace{\left( \ln(4)-1\right)}_{\approx 0.39}\mbox{.}
	\end{align*}}\hspace{1.0cm}\parbox{5cm}{\raggedright\bf RF energy reflected during filling of a superconducting cavity}\hfill
	\parbox[b]{8mm}{\begin{equation}  \end{equation}}

In the same way we can calculate the amount of energy that is reflected during the decay of the electromagnetic fields:\\  
  	\parbox{9cm}{\[
	W_{r, \,decay} = P_{gen} \int^{\infty}_{0} e^{-\frac{t}{\tau_l}} = P_{gen} \cdot \tau_l\mbox{.}
	\]}\hspace{1.0cm}\parbox{5cm}{\raggedright\bf RF energy reflected during in a superconducting cavity}\hfill
	\parbox[b]{8mm}{\begin{equation}  \end{equation}}

If we want to calculate the power needs for the cryogenic system, we need to integrate over the RF power, which is dissipated in the cavity walls during filling of the cavity, beam acceleration, and decay of the fields:\\
  	\parbox{9cm}{\begin{align*}
	W_{d, \,filling} &= \underbrace{\frac{\left(2V_{acc}\right)^2}{\left(R/Q\right)Q_0}}_{P_{d,\,steady\,state}} \int^{\ln(4)\tau_l}_0 \left(1-e^{-\frac{t}{2\tau_l}} \right)^2 dt \\ 
	                    &=  P_{d,\,steady\,state} \cdot \tau_l \underbrace{\left( 8\ln(2)-5\right)}_{\approx 0.55}\mbox{,}
	\end{align*}}\hspace{1.0cm}\parbox{5cm}{\raggedright\bf energy dissipated during filling of a superconducting cavity}\hfill
	\parbox[b]{8mm}{\begin{equation}  \end{equation}}\\	
  	\parbox{9cm}{\begin{align*}
	W_{d, \,decay} &= P_{d,\, steady\,state} \int^{\infty}_{0} e^{-\frac{t}{\tau_l}} \\
	&= P_{d,\,steady\,state} \cdot \tau_l\mbox{.} \hspace{1.5cm}
	\end{align*}}\hspace{1.0cm}\parbox{5cm}{\raggedright\bf energy dissipated during field decay in a superconducting cavity}\hfill
	\parbox[b]{8mm}{\begin{equation}  \end{equation}}

Taking into account the extra time needed to fill the cavities ($\tau_l\ln(4)$) and the power dissipated on the cavity surface during filling and decaying of the fields we can calculate the effective duty cycles for the RF system and the cryogenic power needs by adding up the \\
  	\parbox{9cm}{\begin{align*}
	D_{beam} &= f_{rep}\cdot t_{beam}\mbox{,} \\
	D_{generator} &= f_{rep} \cdot \left( 1.39 \tau_l + t_{beam} \right)\mbox{,} \\
	D_{cryo} &= f_{rep} \cdot \left( 1.55 \tau_l + t_{beam} \right)\mbox{.} \hfill
	\end{align*}}\hspace{0.2cm}\parbox{5cm}{\raggedright \begin{align*} 
	& \mbox{\bf beam duty cycle}\\ 
	& \mbox{\bf generator duty cycle}\\ 
	& \mbox{\bf cryo duty cycle} \end{align*}}\hfill
	\parbox[b]{8mm}{\begin{equation}  \end{equation}}\\	

To demonstrate the impact of filling and decay times on the various duty cycles we use the expected cavity parameters of the SPL \cite{Gerigk10} as given in Table~\ref{tab:SPLnom}
\begin{table}[h!]
\begin{center}
\caption{Expected parameters for the SPL 5-cell $\beta=1$ cavities}
\label{tab:SPLnom}
\begin{tabular}{p{7cm}c}
\hline\hline
\textbf{frequency}  & 704.4\,MHz \\
\textbf{(R/Q)}    & 570\,$\Omega$ \\
$\boldsymbol E_{\mathbf acc}$ & 25\,MV/m \\
\textbf{beam current (average pulse current)} & 40\,mA \\
\textbf{synchronous phase} & $-15^{\circ}$ \\
\textbf{beam pulse length} & 0.4\,ms \\
\textbf{repetition rate} & 50\,Hz \\
\hline\hline
\end{tabular}
\end{center}
\end{table}
and calculate the various duty cycles:\\
  	\parbox{9cm}{\begin{align*}
	\tau_l &= \mbox{0.27\,ms} \\
	t_{fill} &= \mbox{0.38\,ms} \\
	D_{beam} &= \mbox{2\%} \\
	D_{generator} &= \mbox{3.89\%} \\
	D_{cryo} &= \mbox{4.11\%} \hfill
	\end{align*}}\hspace{1.0cm}\parbox{5cm}{\raggedright \begin{align*} 
	&\mbox{\bf filling time constant}\\
	&\mbox{\bf total filling time}^*\\
	&\mbox{\bf beam duty cycle}\\ 
	&\mbox{\bf generator duty cycle}\\ 
	&\mbox{\bf cryo duty cycle} \end{align*}}\hfill
	\parbox[b]{8mm}{\begin{equation}  \end{equation}}\\	
{\small $^*$ assuming a generator power that exactly covers the power needs during the beam passage}

We can see that the filling of the cavities and the subsequent decay of the fields basically double the duty cycles for the RF and the cryogenic system. So despite the fact that the surface losses are negligible compared to the beam power, we are `wasting' $\approx$ 50\% of the RF power and we also have to provide a cryogenic infrastructure to run the cavities (to cool 1\,W at 2\,K we need approximately 750\,W of compressor power). The actual duty cycles are of course highly dependent on the desired cavity voltage and the time structure of the pulses, which means that for each particular application one has to look very carefully at the operational needs in order to decide whether it is more economical to use normal or superconducting cavities. In addition one should consider the increased R\&D time for superconducting cavities and their increased need for inter-cavity space before coming to a decision.



\section*{Acknowledgements}

For the preparation of this lecture I have made extensive use of the material listed below in the Bibliography.

\section*{Bibliography}
\begin{itemize}
\item M. Vretenar, Introduction to RF linear accelerators, CAS CERN Accelerator School: Introduction to Accelerator Physics, Frascati, 2008.
\item T. P. Wangler, Principles of RF Linear Accelerators (Wiley, New York, 1998).
\item D. J. Warner, Fundamentals of electron linacs, CAS CERN Accelerator School: Cyclotrons, Linacs and Their Applications, La Hulpe, Belgium, 1994, S. Turner (Ed.), CERN 96-02, pp. 17--37.
\item H. Padamsee, J. Knobloch, and T. Hays, RF Superconductivity for Accelerators, 2nd Edition (Wiley, Weinheim, 2008).
\end{itemize}




\begin{thebibliography}{99}
\bibitem{Klingbeil} H. Klingbeil, lecture on `Ferrite cavities', CAS RF School 2010, Ebeltoft, Denmark.
\bibitem{JPARCchopper} S. Wang, S. Fu, and T. Kato, The development and beam test of an RF chopper system for J-PARC, Nucl. Instrum. Meth. {\bf 547}  (2005) 302--312.
\bibitem{Fritzchopper} F. Caspers et al., The CERN-SPL chopper concept and final layout, Proc. EPAC 2004, Lucerne, Switzerland. 
\bibitem{CRAB} G. Burt, lecture on `Transverse deflecting cavities', CAS RF School 2010, Ebeltoft, Denmark. 
\bibitem{HBraun} H. Braun, lecture on `Beam monitoring with RF cavities', CAS RF School 2010, Ebeltoft, Denmark.
\bibitem{CASMaurizio} M. Vretenar, `Low-beta structures', CAS RF School 2010, Ebeltoft, Denmark.
\bibitem{Knapp} D. E. Nagle, and E. A. Knapp, B. C. Knapp, Coupled resonator model for standing wave accelerator tanks,  Rev. Sci. Instrum. {\bf 38} (1967) 1583.
\bibitem{Schriber01} S. Schriber, Characteristics of full-cell terminated RF structures: results of analogue studies, CERN/PS/2001-067 (PP).
\bibitem{Schriber2} S. Schriber, Analog analysis of pi-mode structures: results and implications, Phys. Rev. Spec. Top. Accel. Beams, {\bf 4} (2001) 122001. 
\bibitem{clic} http://www.cern.ch/clic-study
\bibitem{ilc} http://www.linearcollider.org
\bibitem{Miller86} R. H. Miller, Comparison of standing wave and travelling wave structures, LINAC86.
\bibitem{Moiseev00} V. A. Moiseev, V. V. Paramonov, and K. Floettmann, Comparison of standing and travelling wave operations for positron pre-accelerator in the TESLA Linear Collider, EPAC 2000.
\bibitem{CASErk} E. Jensen, `Cavity Basics', CAS RF School 2010, Ebeltoft, Denmark.
\bibitem{Tiede08} R. Tiede, U. Ratzinger, H. Podlech, C. Zhang, and G. Clemente, {\sl talk on:} Konus beam dynamics using H-mode cavities, HB 2008.
\bibitem{ClementeThesis} G. Clemente, Room temperature CH-DTL and its application for the FAIR proton injector, PhD thesis, University of Frankfurt, 2007. 
\bibitem{Ciprian08} C. Plostinar, Comparative assessment of HIPPI normal-conducting structures, CARE-report-08-071-HIPPI.
\bibitem{RatzingerHabil} U. Ratzinger, Effiziente Hochfrequenz-Linearbeschleuniger f\"{u}r leichte und schwere Ionen, Habilitationsschrift, July 1998, University of Frankfurt. 
\bibitem{Linac3} J. Broere, H. Kugler, M. Vretenar, U. Ratzinger, and B. Krietenstein, High power conditioning of the 202\,MHz IH tank 2 at the CERN Linac3, LINAC 1998.
\bibitem{Podlech99} H. Podlech, M. Grieser, R. von Hahn, R. Repnow, and D. Schwalm, The 7-gap-resonator-accelerator for the REX-ISOLDE Linac, PAC 1999.
\bibitem{HFitze}
H. Fitze et al., Developments at PSI (including new RF cavity), CYC2004.
\bibitem{Morvillo} M. Morvillo et al., The CERN PS 13.3 - 20\,MHz RF system for LHC, PAC 2003. 
\bibitem{Delayen88} J. R. Delayen, Superconducting accelerating structures for high-current ion beams, LINAC 1988.
\bibitem{Zaplatin} E. Zaplatin et al., Triple spoke cavities at FZJ, EPAC 2004.
\bibitem{Kilpatrick} W. D. Kilpatrick, Criterion for vacuum sparking designed to include both RF and DC, Rev. Sci. Instrum. {\bf 28} (1957) 824.
\bibitem{Gerigk06}
F. Gerigk, Formulae to calculate the power consumption of the SPL SC cavities, CERN-AB-2006-011, 2006.
\bibitem{Gerigk10} F. Gerigk et al., Layout and machine optimisation for the SPL at CERN, LINAC 2010. 
\end{thebibliography}
\end{document}